\pdfoutput=1
\documentclass[prx,twocolumn,showpacs,amsmath,amssymb,floatfix]{revtex4-1}

\usepackage{graphicx}  
\usepackage{dcolumn}   
\usepackage{bm}        
\usepackage{amssymb}   
\usepackage{amsmath}
\usepackage{mathrsfs}
\usepackage{epigraph}
\usepackage{braket}
\usepackage{gensymb}
\usepackage{lmodern}
\usepackage{lipsum}
\usepackage{marvosym}
\usepackage{ tipa }
\usepackage{bbold}
\usepackage{dsfont}
\usepackage{esint}
\usepackage{xcolor}
\usepackage{appendix}
\usepackage{hyperref}
\usepackage{natbib}

 \DeclareMathOperator{\Tr}{Tr}
 \DeclareMathOperator{\DM}{\boldsymbol{\rho}}

\DeclareMathOperator{\deff}{\equiv}
\DeclareMathOperator{\id}{\bold{\mathds{1}}}

\begin{document}

\title{Time-retarded damping and magnetic inertia in the Landau-Lifshitz-Gilbert equation self-consistently coupled to electronic time-dependent nonequilibrium Green functions}

\author{Utkarsh Bajpai}
\affiliation{Department of Physics and Astronomy, University of Delaware, Newark, DE 19716, USA}
\author{Branislav K. Nikoli\'{c}}
\email{bnikolic@udel.edu}
\affiliation{Department of Physics and Astronomy, University of Delaware, Newark, DE 19716, USA}
		
\begin{abstract}
The  conventional Landau-Lifshitz-Gilbert (LLG) equation is a widely used tool to describe dynamics of local magnetic moments, viewed as classical vectors of fixed length, with their change assumed to take place simultaneously with the cause. Here we demonstrate that recently developed [M. D. Petrovi\'{c} {\em et al.}, Phys. Rev. Applied {\bf 10}, 054038 (2018)]  self-consistent coupling of the LLG equation to time-dependent quantum-mechanical description of electrons---where nonequilibrium spin density from time-dependent nonequilibrium Green function (TDNEGF) calculations is inserted within a torque term into the LLG equation while local magnetic moments evolved by the LLG equation introduce time-dependent potential in the quantum Hamiltonian of electrons---microscopically generates time-retarded damping in the LLG equation described by a memory kernel which is also spatially dependent. For sufficiently slow dynamics of local magnetic moments on the memory time scale, the kernel can be expanded into power series to extract the Gilbert damping (proportional to first time derivative of magnetization) and magnetic inertia (proportional to second time derivative of magnetization) terms whose parameters, however, are {\em time-dependent} in contrast to time-independent parameters used in the conventional LLG equation. We use examples of single or multiple local magnetic moments precessing in an external magnetic field, as well as field-driven motion of a magnetic domain wall (DW), to quantify the difference in their time evolution computed  from conventional  LLG equation vs. TDNEGF+LLG quantum-classical hybrid approach. The faster DW motion predicted by TDNEGF+LLG approach reveals that important quantum effects, stemming essentially from a finite amount of time which it takes for conduction electron spin to react to the motion of classical local magnetic moments, are missing from conventional classical micromagnetics simulations. We also demonstrate large discrepancy between TDNEGF+LLG-computed numerically exact and, therefore, nonperturbative result for charge current pumped by a moving DW and the same quantity computed by perturbative spin motive force formula combined with the conventional LLG equation.  
\end{abstract}

\maketitle

%!!!!!!!!!!!!!!!!!!!!!!!!!!!!!!!!!!!!!!!!!!!!!!!!!!!!!!!!!!!!!!!!!!!!!!!!!!!!!!!!!!!!!!!!!!!!!!!!!!!!!!!!!!!!!!!!!!!!!!!!!!!!!!!!!!!!!!!!!!!!!!!!!!!!!!!!!!!!!!!!!!!!!!!!!!!!!!!
\section{Introduction}\label{sec:intro}
%!!!!!!!!!!!!!!!!!!!!!!!!!!!!!!!!!!!!!!!!!!!!!!!!!!!!!!!!!!!!!!!!!!!!!!!!!!!!!!!!!!!!!!!!!!!!!!!!!!!!!!!!!!!!!!!!!!!!!!!!!!!!!!!!!!!!!!!!!!!!!!!!!!!!!!!!!!!!!!!!!!!!!!!!!!!!!!!

The conventional Landau-Lifshitz-Gilbert (LLG) equation~\cite{Bertotti2009,Wieser2013,Wieser2015} is the cornerstone of numerical micromagnetics~\cite{Kumar2017} and atomistic spin dynamics~\cite{Evans2014} where one simulates the classical time evolution of many magnetic units coupled by exchange or magnetostatic interactions. The LLG equation  
\begin{equation}\label{eq:llg}
\frac{\partial \mathbf{m}(\mathbf{r},t)}{\partial t} = - g \mathbf{m}(\mathbf{r},t) \times \mathbf{B}_\mathrm{eff}(\mathbf{r},t) + \lambda_\mathrm{G} \mathbf{m}(\mathbf{r},t) \times \frac{\partial \mathbf{m}(\mathbf{r},t)}{\partial t},
\end{equation}
describes time evolution of $\mathbf{m}(\mathbf{r},t)$ as the unit vector $|\mathbf{m}|=1$ of constant length representing the direction of the local magnetization. Here $g$ is the gyromagnetic ratio and $\mathbf{B}_\mathrm{eff}$ is the sum of an external magnetic field and effective magnetic fields due to magnetic anisotropy and exchange coupling (additional stochastic magnetic field can contribute to $\mathbf{B}_\mathrm{eff}$ to take into account finite temperature effects~\cite{Nunez2008}). The second term on the right-hand side of Eq.~\eqref{eq:llg} is introduced phenomenologically to break the time-inversion symmetry, thereby generating a damping mechanism. The conventional intrinsic Gilbert damping  $\lambda_\mathrm{G}$ is assumed to be materials specific and, therefore, time-independent parameter. It is typically computed using the so-called breathing Fermi surface~\cite{Kambersky2007} or torque-torque correlation formulas~\cite{Gilmore2007} within single-particle quantum-mechanical framework (additional many-body processes have to be taken into account to make $\lambda_\mathrm{G}$ finite in the clean limit at low temperatures~\cite{Mahfouzi2017a}). In the original form, Eq.~\eqref{eq:llg}  is written for a  bulk material as a highly nonlinear partial differential equation. It can also be re-written for a macrospin or a lattice of atomic spins leading to a system of nonlinear ordinary differential equations~\cite{Evans2014}. 

In the case of conducting ferromagnets, LLG equation has to be extended by including additional terms, such as: ({\em i}) spin-transfer torque~\cite{Ralph2008}  $\mathbf{T} \propto \langle \hat{\mathbf{s}} \rangle \times \mathbf{m}$ due to injected electrons generating nonequilibrium spin density  $\langle \hat{\mathbf{s}} \rangle$ that is noncollinear to local magnetization; ({\em ii}) additional Gilbert damping, $(g_{\uparrow\downarrow}/4\pi) \mathbf{m} \times \partial \mathbf{m}/\partial t$, due to pumping of spin currents by the dynamics of $\mathbf{m}(t)$ where $g_{\uparrow\downarrow}$ is the so-called spin-mixing conductance~\cite{Tserkovnyak2005}; ({\em iii}) additional {\em nonlocal} Gilbert  damping~\cite{Zhang2009b,Wong2009,Tserkovnyak2009,Kim2012b,Nembach2013,Weindler2014,Li2016a}, \mbox{$\mathbf{m} \times (\mathcal{D} \cdot \partial \mathbf{m}/\partial t)$}, due to spin pumping by noncollinear magnetic textures where \mbox{$\mathcal{D}_{\alpha \beta} = \eta \sum_i (\mathbf{m} \times \partial_i \mathbf{m})_\alpha (\mathbf{m} \times \partial_i \mathbf{m})_\beta$} is the $3 \times 3$ damping tensor, $\partial_i=\partial/\partial_i$ and $\alpha,\beta,i \in \{x,y,z \}$; and ({\em iv}) magnetic inertia~\cite{Faehnle2011,Bhattacharjee2012,Kikuchi2015,Sayad2016a,Mondal2017,Mondal2018,Li2015b},  $I \mathbf{m} \times \partial^2 \mathbf{m}/\partial t^2$, of relevance to ultrafast magnetization dynamics. Like the original Gilbert damping parameter $\lambda_\mathrm{G}$ in Eq.~\eqref{eq:llg}, $\mathbf{T}$, $g_{\uparrow\downarrow}$, $\mathcal{D}$ and $I$ require  microscopic quantum-mechanical calculations which are often combined~\cite{Gilmore2007,Mahfouzi2017a,Carva2007,Liu2014a,Wang2008b,Ellis2017,Nikolic2018,Thonig2017} with first-principles Hamiltonians of realistic materials. 

Furthermore, generalizations of LLG equation have been considered to take into account the retardation effects~\cite{Bose2011,Thonig2015}
\begin{eqnarray}\label{eq:memoryllg}
\frac{\partial \mathbf{m}(\mathbf{r},t)}{\partial t} & = &  \int_0^t dt' \int d^3 \mathbf{r}' \Gamma(\mathbf{r},t;\mathbf{r}',t') \mathbf{m}(\mathbf{r}',t') \nonumber \\
&& \times \left[ -g \mathbf{B}_\mathrm{eff}(\mathbf{r}',t') + \lambda_\mathrm{G} \frac{\partial \mathbf{m}(\mathbf{r}',t')}{\partial t'} \right],
\end{eqnarray}
by introducing a memory kernel $\Gamma(\mathbf{r},t;\mathbf{r}',t')$. The memory kernel models space-time correlation between local magnetic moments, i.e., the fact that the cause for the change of local magnetization occurs at time $t - t'$ and at position $\mathbf{r} - \mathbf{r}'$ while the effect at position $\mathbf{r}$ is visible at the later time $t$. It has been specified phenomenologically, such as the sum of an instantaneous and time-dependent part which exponentially decays on a characteristic time scale defining the strength of memory~\cite{Bose2011,Thonig2015}. It is also often simplified~\cite{Thonig2015} by considering time-retardation only, $\Gamma(\mathbf{r},t;\mathbf{r}',t') \rightarrow \Gamma(t,t')$, so that any space-retardation effects are included only through the effective field $\mathbf{B}_\mathrm{eff}(\mathbf{r},t')$. 

The time-retardation described by $\Gamma(t,t')$ is a damping mechanism in addition to well-established mechanisms---the combined effects of spin-orbit coupling and electron-phonon interaction~\cite{Kambersky2007,Gilmore2007}---which govern $\lambda_\mathrm{G}$ in Eq.~\eqref{eq:llg}. However, the magnitude of $\Gamma(t,t')$ cannot be deduced from purely phenomenological considerations~\cite{Bose2011,Thonig2015}. Instead, the introduction of the memory kernel $\Gamma(t,t')$ can be justified microscopically~\cite{Onoda2006d,Nunez2008,Sayad2015,Sayad2016a,Hammar2016,Hammar2017} by using {\em quantum-classical hybrid} approaches, where time-dependent quantum formalism is used to compute  $\langle \hat{\mathbf{s}} \rangle (t)$ which is then fed into the LLG equation, while in turn, local magnetization from the LLG equation generates time-dependent field in the quantum Hamiltonian of electrons. Although electron dynamics is assumed to be much faster than that of local magnetic moments, it still takes finite time for electron spin to react to new position of $\mathbf{m}(\mathbf{r},t)$. This is the fundamental reason for time-retarded damping effects encoded by Eq.~\eqref{eq:memoryllg}, which are present even if the intrinsic Gilbert damping in Eq.~\eqref{eq:llg} is vanishingly small due to small spin-orbit coupling (nonzero $\lambda_\mathrm{G}$ requires spin-orbit coupling~\cite{Kambersky2007,Gilmore2007,Mahfouzi2017a} and scales quadratically with it~\cite{He2013a}). Since classical micromagnetics simulations typically use only the conventional intrinsic Gilbert damping term in Eq.~\eqref{eq:llg}, while not considering explicitly the flow of conduction electrons in the presence of magnetization dynamics, the question arises about the magnitude of  neglected effects like time-retarded damping in standard simulations of magnetic-field- or current-driven dynamics of noncollinear magnetic textures such as magnetic domain walls (DWs)~\cite{Lee2004a,Stiles2007,Li2004,Li2004a,Thiaville2005,Thiaville2007,Martinez2009,Boone2010,Chureemart2011} and skyrmions~\cite{Iwasaki2013a,Sampaio2013}.

%=====FIGURE======
\begin{figure}[h]
\includegraphics[scale=0.07]{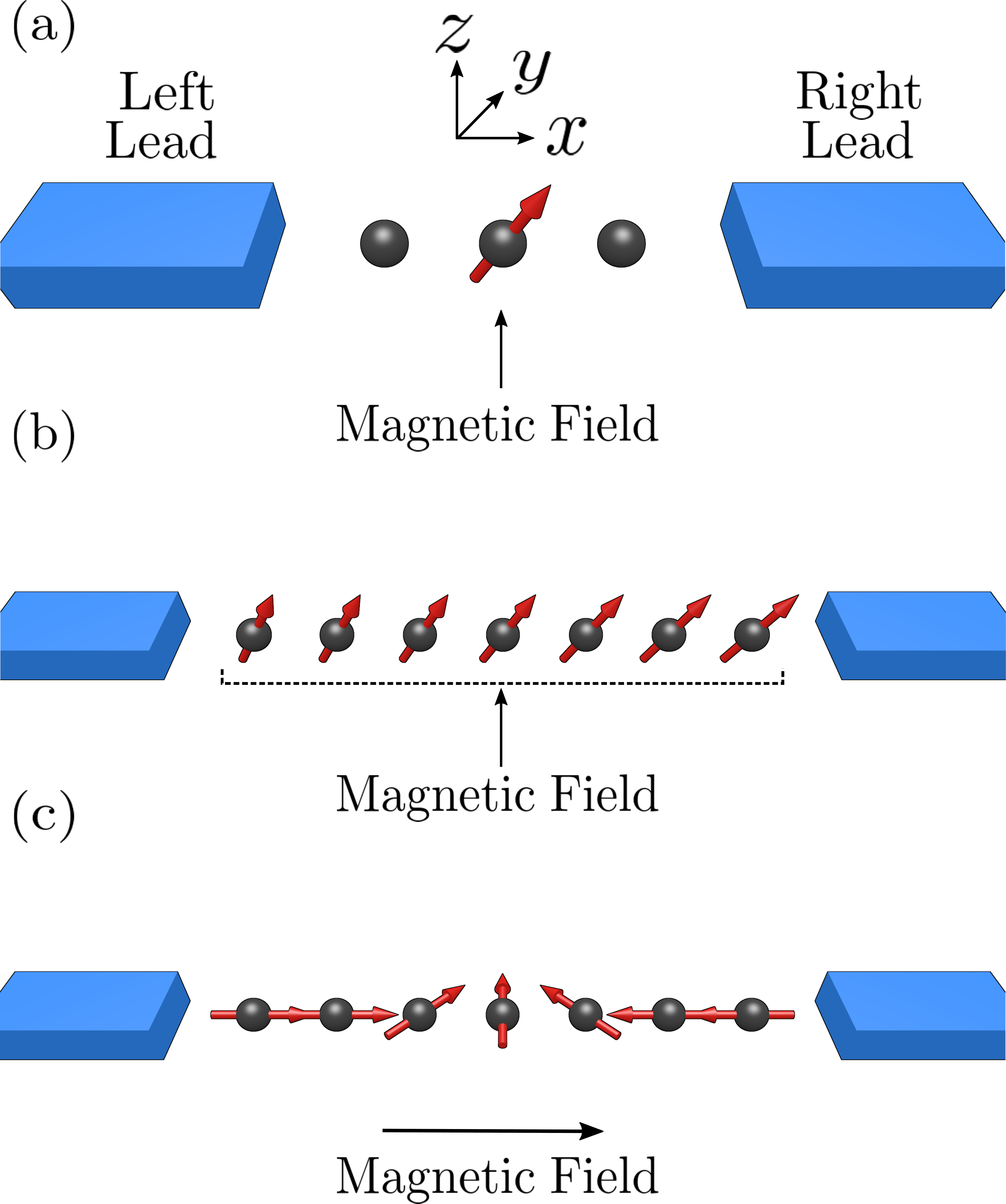}
\caption{Schematic view of two-terminal devices where an infinite 1D TB chain, describing electrons quantum-mechanically, is attached to two macroscopic reservoirs while its middle part hosts: (a) single local magnetic moment, initially oriented in the $+x$-direction, placed in an external magnetic field pointing along the $+z$-direction; (b) 11 local magnetic moments (illustration shows 7 of them), initially oriented in the $+x$-direction, placed in an external magnetic field pointing along the $+z$-direction; (c) three-site-wide head-to-head magnetic DW whose motion is driven by an external magnetic field pointing in the $+x$-direction. Electrons within 1D TB chain and classical local magnetic moments interact via the $s$-$d$ exchange coupling of strength $J_{sd}$, and classical local magnetic moments within the DW in (c) additionally interact with each other via the Heisenberg exchange coupling of strength $J$. }
\label{fig:fig1}
\end{figure}

Although quantum-classical approaches which automatically include time-retardation effects have been discussed previously~\cite{Onoda2006d,Nunez2008,Sayad2015,Sayad2016a,Hammar2016,Hammar2017}, they have been focused on the simples examples where one or two local magnetic moments (pertinent to, e.g., magnetic molecules) interact with either closed electronic quantum system~\cite{Sayad2015,Sayad2016a} (i.e., not attached to macroscopic reservoirs to allow electron spin and charge currents to flow into and from an external circuit), or open electronic quantum system but employing approximations~\cite{Onoda2006d,Nunez2008,Hammar2016,Hammar2017} to obtain  analytical solution. Thus, these approaches are not suitable for simulations of spintronic devices containing large number of noncollinear local magnetic moments.

Here we employ recently developed~\cite{Petrovic2018} \textit{numerically exact and, therefore, nonperturbative} algorithm combining time-dependent nonequilibrium Green function formalism~\cite{Stefanucci2013,Gaury2014}  with the conventional LLG Eq.~\eqref{eq:llg} (TDNEGF+LLG)  to demonstrate how it  effectively generates  time-retardation effects, whose memory kernel can be explicitly extracted in terms of TDNEGFs only in some limits (such as weak electron-spin/local-magnetic-moment interaction and weak coupling of the active region to macroscopic reservoirs). The paper is organized as follows. Section~\ref{sec:hamiltonians} introduces model quantum Hamiltonian for electronic subsystem and classical Hamiltonian for the subsystem comprised of local magnetic moments. In Sec.~\ref{sec:dampingsec}, we show how the nonequilibrium expectation value of spin density
\begin{equation}\label{eq:spin}
\langle \hat{\mathbf{s}} \rangle^i (t) = \frac{\hbar}{2} \mathrm{Tr} \,[(\boldsymbol{\rho}_\mathrm{neq}(t)-\boldsymbol{\rho}_\mathrm{eq})\ket{i}\bra{i}\otimes {\bm{\sigma}}],   
\end{equation}
inserted into the LLG Eq.~\eqref{eq:llg} generates a memory kernel because of the structure of the nonequilibrium time-dependent density matrix $\boldsymbol{\rho}_\mathrm{neq}(t)$ obtained from TDNEGF calculations. Here $\boldsymbol{\rho}_\mathrm{eq}$ is the grand canonical equilibrium density matrix; ${\bm \sigma}=(\hat{\sigma}^x,\hat{\sigma}^y,\hat{\sigma}^z)$ is the vector of the Pauli matrices; $\ket{i}$ electron orbital centered on site $i$; and the operator $\ket{i}\bra{i}\otimes{\bm \sigma}$ acts in the composite Hilbert space $\mathscr{H} = \mathscr{H}_\mathrm{orb}\otimes\mathscr{H}_\mathrm{spin}$ of electronic orbital and spin degrees of freedom. In this Section, we also discuss how in the limit of slow magnetization dynamics the memory kernel can be expanded in a Taylor series in order to extract conventional Gilbert damping and magnetic inertia terms, but with time- and spatially-dependent  parameters $\lambda^\mathrm{D}_i(t)$ and $I^\mathrm{D}_i(t)$. In Secs.~\ref{sec:singlespin}--\ref{sec:magneticdw} we compare the dynamics of local magnetic moments driven by an external magnetic field as computed by TDNEGF+LLG vs. conventional LLG simulations for three one-dimensional (1D) examples depicted in Fig.~\ref{fig:fig1}(a)--(c), respectively. Sec.~\ref{sec:magneticdw} also compares pumped charge current due to the DW motion as computed by TDNEGF+LLG vs. the widely-used spin motive force (SMF) theory \cite{Barnes2007,Zhang2009b} combined~\cite{Ohe2009,Shimada2015,Yamane2016} with the conventional LLG equation. We conclude in Sec.~\ref{sec:conclusions}.
%
%!!!!!!!!!!!!!!!!!!!!!!!!!!!!!!!!!!!!!!!!!!!!!!!!!!!!!!!!!!!!!!!!!!!!!!!!!!!!!!!!!!!!!!!!!!!!!!!!!!!!!!!!!!!!!!!!!!!!!!!!!!!!!!!!!!!!!!!!!!!!!!!!!!!!!!!!!!!!!!!!!!!!!!!!!!!!!!!
 \section{Models and Methods}\label{sec:models}
 \subsection{Coupled quantum and classical Hamiltonians}\label{sec:hamiltonians}
%!!!!!!!!!!!!!!!!!!!!!!!!!!!!!!!!!!!!!!!!!!!!!!!!!!!!!!!!!!!!!!!!!!!!!!!!!!!!!!!!!!!!!!!!!!!!!!!!!!!!!!!!!!!!!!!!!!!!!!!!!!!!!!!!!!!!!!!!!!!!!!!!!!!!!!!!!!!!!!!!!!!!!!!!!!!!!!!
The conduction electron subsystem is modeled by a quantum Hamiltonian %
%=====EQUATION======
\begin{equation} \label{eq:ham}
\bold{H}(t) = -\gamma\sum_{\braket{ij}}\hat{c}_{i}^\dagger\hat{c}_i - J_{sd} \sum_{i}\hat{c}_i^\dagger\boldsymbol{\sigma} \cdot \bold{M}_i(t) \hat{c}_i,
\end{equation}
which is (assumed to be 1D for simplicity) tight-binding (TB) model where electron interacts with magnetic moments localized at sites $i$ and described by the classical vector $\bold{M}_i(t)$ of unit length. Here $\hat{c}_i^\dagger = (\hat{c}_{i\uparrow}^\dagger,\hat{c}_{i\downarrow}^\dagger)$  is a row vector containing operators $\hat{c}_{i\sigma}^\dagger$ which create an electron of spin $\sigma=\uparrow,\downarrow$ at site $i$; $\hat{c}_i$ is a column vector that contains the corresponding  annihilation operators; \mbox{$\gamma=1$ eV} is the nearest neighbor hopping; and  $J_{sd}$ is the $s$-$d$ exchange coupling parameter between conduction electrons and local magnetic moments. The active region of devices depicted in Fig.~\ref{fig:fig1}(a)--(c) consists of  1, 11 and 21 TB sites, respectively. These are attached to the left (L) and right (R) semi-infinite ideal leads modeled by the same Hamiltonian in Eq.~\eqref{eq:ham} but with \mbox{$J_{sd}=0$ eV}. The leads are assumed to terminate into macroscopic reservoirs kept at the same chemical potential since we do not apply any bias voltage to the devices in Fig.~\ref{fig:fig1}(a)--(c).

The classical Hamiltonian describing the local magnetic moments is given by 
%=====EQUATION======
\begin{multline}  \label{eq:classmm}
\mathcal{H} = -J\sum_{ij} \bold{M}_i \cdot \bold{M}_j - \mu_M \sum_{i}\bold{M}_i \cdot \bold{B}_{\mathrm{ext}}^{i} - K\sum_{i}(\mathrm{M}_i^x)^2 \\ -J_{sd}\sum_i\braket{\boldsymbol{\hat{\mathrm{s}}}}^i \cdot \bold{M}_i,
\end{multline}
where $J$ is the Heisenberg exchange coupling parameter; $\bold{B}^i_{\mathrm{ext}}$ is the applied external magnetic field; $K$ is the magnetic anisotropy (in the $x$-direction) and $\braket{\hat{\mathbf{s}}}^i$ is the nonequilibrium electronic spin density computed from Eq.~\eqref{eq:spin}. 

%!!!!!!!!!!!!!!!!!!!!!!!!!!!!!!!!!!!!!!!!!!!!!!!!!!!!!!!!!!!!!!!!!!!!!!!!!!!!!!!!!!!!!!!!!!!!!!!!!!!!!!!!!!!!!!!!!!!!!!!!!!!!!!!!!!!!!!!!!!!!!!!!!!!!!!!!!!!!!!!!!!!!!!!!!!!!!!!
\subsection{Time-retarded damping and magnetic inertia in the LLG equation self-consistently coupled to TDNEGF }\label{sec:dampingsec}
%!!!!!!!!!!!!!!!!!!!!!!!!!!!!!!!!!!!!!!!!!!!!!!!!!!!!!!!!!!!!!!!!!!!!!!!!!!!!!!!!!!!!!!!!!!!!!!!!!!!!!!!!!!!!!!!!!!!!!!!!!!!!!!!!!!!!!!!!!!!!!!!!!!!!!!!!!!!!!!!!!!!!!!!!!!!!!!!
The quantum equation of motion for the nonequilibrium density matrix of electrons~\cite{Croy2009,Popescu2016}
%=====EQUATION======
\begin{equation}  \label{eq:dme}
i\hbar\frac{\partial \DM_\mathrm{neq}(t)}{\partial t} = [\bold{H}(t),\DM_\mathrm{neq}(t)] + \sum_{p=\mathrm{L,R}} i [\bold{\Pi}_{p}(t) + \bold{\Pi}_{p}^\dagger(t) ] .
\end{equation}
is an example of a master equation for an open (i.e., connected to macroscopic reservoirs) quantum system~\cite{Breuer2002} due to the presence of the second term on the right hand side, in addition to standard terms of the von Neumann equation. This term and the density matrix itself can be expressed using TDNEGF formalism~\cite{Stefanucci2013,Gaury2014} as  
\begin{equation}\label{eq:neqdm}
{\bm \rho}_\mathrm{neq}(t) = \frac{1}{i}\mathbf{G}^<(t,t') |_{t=t'},
\end{equation}
%
%=====EQUATION======
\begin{multline} \label{eq:currm}
\bold{\Pi}_{p}(t') = \int_{-\infty}^{t'} dt_1 [\bold{G}^>(t',t_1)\bold{\Sigma}^<_{p}(t_1,t') - \\ \bold{G}^<(t',t_1)\bold{\Sigma}_{p}^>(t_1,t')].
\end{multline}
The central quantities of the TDNEGF formalism are the retarded $G_{ii'}^{r,\sigma\sigma'}(t,t') = -i\Theta(t-t')\braket{\{ \hat{c}_{i\sigma}(t),\hat{c}_{i'\sigma'}(t)\}}$ and the lesser $G^{<,\sigma \sigma'}_{ii'}(t,t') = i\braket{\hat{c}_{i'\sigma'}^\dagger(t')\hat{c}_{i\sigma}(t)}$ Green functions (GFs) which describe the available density of states and how electrons occupy those states, respectively. In addition, it is also useful to introduce the greater GF, $\bold{G}^{>}(t,t') = [\bold{G}^{<}(t',t)]^{\dagger}$, and the advanced GF, $\bold{G}^a(t,t') = [\bold{G}^r(t,t')]^{\dagger}$. The current matrices $\bold{\Pi}_p(t)$ make it possible to compute directly~\cite{Croy2009,Popescu2016} charge current 
\begin{equation}\label{eq:c_curr}
I_{p}(t) = \frac{e}{\hbar}\Tr[\bold{\Pi}_p(t)],
\end{equation}
and spin current 
\begin{equation}\label{eq:spin_curr}
I_{p}^{S_{\alpha}}(t) = \frac{e}{\hbar}\Tr[\hat{\sigma}^\alpha \bold{\Pi}_p(t)],
\end{equation}
in the L and R semi-infinite leads. The equation of motion for the lesser and greater GFs is given by
%=====EQUATION======
\begin{multline} \label{eq:gfe}
i\hbar\frac{\partial \bold{G}^{>,<}(t,t_1)}{\partial t} = \bold{H}(t)\bold{G}^{>,<}(t,t_1) + \\ \int\limits_{-\infty}^{+\infty} dt_2 \bigg[ \bold{\Sigma}^r_{\mathrm{tot}}(t,t_2)\bold{G}^{>,<}(t_2,t) + \bold{\Sigma}^{>,<}_{\mathrm{tot}}(t,t_2)\bold{G}^a(t_2,t)\bigg],
\end{multline}
where $\bold{\Sigma}^{r,>,<}_{\mathrm{tot}}(t,t_2) = \sum_{p=\mathrm{L,R}}\bold{\Sigma}_{p}^{r,>,<}(t,t_2)$ and $\bold{\Sigma}_{p}^{r,>,<}(t,t_2)$ are the lead self-energy matrices~\cite{Gaury2014,Croy2009,Popescu2016}.

The classical equation of motion for the magnetic moment localized at site $i$ is the Landau-Lifshitz equation
 %=====EQUATION======
\begin{equation} \label{eq:llg_sd}
\frac{\partial \bold{M}_i(t)}{\partial t} = -g\bold{M}_i(t)\times \bold{B}^{\mathrm{eff}}_i(t),
\end{equation}
where the effective magnetic field is \mbox{$\bold{B}^{\mathrm{eff}}_i(t) = -\frac{1}{\mu_M}\partial \mathcal{H}/\partial \bold{M}_i$} and $\mu_M$ is the magnitude of the magnetic moment~\cite{Evans2014}. 

The full TDNEGF+LLG framework~\cite{Petrovic2018}, which we also denote as TDNEGF $\leftrightarrows$ LLG, consists of self-consistent combination of Eq.\eqref{eq:dme} and \eqref{eq:llg_sd} where one first solves for the nonequilibrium electronic spin density in Eq.\eqref{eq:spin}, which is then fed into Eq. \eqref{eq:llg_sd} to propagate local magnetic moments $\bold{M}_i(t)$ in the next time step. Evolving  ${\bm \rho}_\mathrm{neq}(t)$ via Eq.~\eqref{eq:dme} requires time step \mbox{$\delta t=0.1$ fs} for numerical stability, and we use the same time step to evolve LLG or Landau-Lifshitz equations for $\bold{M}_i(t)$. These updated local magnetic moments are fed back into the quantum Hamiltonian of conduction electron subsystem in Eq. \eqref{eq:dme}. Thus obtained solutions for $\bold{M}_i(t)$, $\langle \hat{\mathbf{s}} \rangle^i (t)$, $I_{p}(t)$ and $I_p^{S_{\alpha}}(t)$ are numerically exact. For testing the importance of the self-consistent feedback loop, we also use TDNEGF $\leftarrow$ LLG where TDNEGF is utilized to obtain $I_{p}(t)$ and $I_p^{S_{\alpha}}(t)$ while the local magnetic moments are evolved solely by the conventional LLG Eq.~\eqref{eq:llg}, i.e., by using $J_{sd} \equiv 0$ in Eq.~\eqref{eq:classmm} but $J_{sd} \neq 0$ is used in Eq.~\eqref{eq:ham}.

In the weak-coupling limit~\cite{Onoda2006d,Zhang2004a} (i.e., small $J_{sd}$) for electron-spin/local-magnetic-moment interaction it is possible to extract explicitly the generalized LLG equation with a  memory kernel. For this purpose we use the following expansions in the powers of small $J_{sd}$
%=====EQUATION======
\begin{equation} \label{eq:dm_exp}
\bold{\DM}_\mathrm{neq}(t) = \sum_{n=0}^{\infty}\bold{\DM}_n(t)J_{sd}^n,
\end{equation}
%
%=====EQUATION======
\begin{equation} \label{eq:pi_exp}
\bold{\Pi}_{p}(t') = \sum_{n=0}^{\infty}\bold{\Pi}_{p}^{(n)}(t')J_{sd}^n,
\end{equation}
%
%=====EQUATION======
\begin{equation} \label{eq:gf_exp}
\bold{G}^{r,a,>,<}(t',t_1) = \sum_{n=0}^{\infty}\bold{G}^{r,a,>,<}_n(t',t_1)J_{sd}^n.
\end{equation}
In Appendix~\ref{sec:app},  we show how to combine Eqs.\eqref{eq:dme}, \eqref{eq:gfe}, \eqref{eq:dm_exp}, \eqref{eq:pi_exp} and \eqref{eq:gf_exp} to obtain the perturbative equation 
%=====EQUATION======
\begin{multline} \label{eq:llg_mem}
\frac{\partial \bold{M}_i(t)}{\partial t} = -g\bigg[\bold{M}_i(t)\times \bold{B}^{\mathrm{eff},0}_{i}(t) + \\  \frac{J^2_{sd}}{\mu_M}\sum_{p=\mathrm{L,R}} \bold{M}_i(t) \times  \int \limits_{-\infty}^{+\infty} dt'' \bold{M}_i(t'') \{\mathrm{K}^{p}_i(t'',t) + \mathrm{K}^{p*}_i(t'',t)\} \bigg],
\end{multline}
for the dynamics of each local magnetic moment at site $i$, by retaining only the terms linear in $J_{sd}$ in Eqs.~\eqref{eq:dm_exp}--\eqref{eq:gf_exp}. Here $\bold{B}^{\mathrm{eff},0}_{i} \deff -\frac{1}{\mu_M}\partial \mathcal{H}^{0}/\partial \bold{M}_i$, $\mathcal{H}^0$ is the classical Hamiltonian in Eq.~\eqref{eq:classmm} with $J_{sd} \equiv 0$ and $\mathrm{K}^{p}_i(t'',t)$ is defined in Appendix~\ref{sec:app}. The physical origin~\cite{Sayad2015} of time-retardation effects described by the second term in Eq.~\eqref{eq:llg_mem}  is that, even though electron dynamics is much faster than the dynamics of local magnetic moments, the nonequilibrium spin density in Eq.~\eqref{eq:spin} is always behind $\bold{M}_i(t)$ and, therefore, never parallel to it which introduces spin torque term into the Landau-Lifshitz Eq.~\eqref{eq:llg_sd}. In other words it takes finite amount of time for conduction electron spin to react to the motion of classical local magnetic moments, so that nonequilibrium electrons effectively mediate interaction of $\bold{M}_i(t)$ with the same local magnetic moment at time $t'<t$. In the full TDNEGF+LLG, such retardation effects are mediated by the nonequilibrium electrons starting at site $i$ at time $t'$ and returning back to the same site at time $t>t'$, while in the perturbative limit the same effect is captured by the second term in Eq.~\eqref{eq:llg_mem}. The perturbative formula Eq.~\eqref{eq:llg_mem} is expected~\cite{Sayad2015} to breakdown after propagation over time $t \sim \hbar/J_{sd}$.

Further approximation to Eq.~\eqref{eq:llg_mem} can be made by considering sufficiently slow dynamics of local magnetic moments so that higher order terms in the Taylor series  
%=====EQUATION======
\begin{equation}\label{eq:spin_expan}
\bold{M}_i(t'') \approx \bold{M}_i(t) + \frac{\partial \bold{M}_i(t)}{\partial t}(t''-t) +  \frac{1}{2}\frac{\partial^2 \bold{M}_i(t)}{\partial t^2}(t''-t)^2 + \ldots,
\end{equation}
can be neglected. By defining the following quantities
%=====EQUATION======
\begin{equation}\label{eq:1stdev}
\lambda^\mathrm{D}_{p,i}(t) \equiv \int \limits_{-\infty}^{+\infty}dt'' (t''-t)[\mathrm{K}^{p}_i(t'',t) + \mathrm{K}^{*p}_i(t'',t)],
\end{equation}
and 
\begin{equation}\label{eq:2nddev}
I^\mathrm{D}_{p,i}(t) \equiv \frac{1}{2}\int \limits_{-\infty}^{+\infty}dt'' (t''-t)^2[\mathrm{K}^{p}_i(t'',t) + \mathrm{K}^{*p}_i(t'',t)],
\end{equation}
and by retaining terms up to the second order in Eq.~\eqref{eq:spin_expan} we obtain the conventionally looking LLG equation
%=====EQUATION======
\begin{multline} \label{eq:llg_nu}
\frac{\partial \bold{M}_i(t)}{\partial t} = -g\bigg[\bold{M}_i(t)\times \bold{B}^{\mathrm{eff},0}_i(t) + \\ \frac{J^2_{sd}}{\mu_M} \bigg\{\sum_{p=\mathrm{L,R}}\lambda_{p,n}^\mathrm{D}(t)\bigg\}\bold{M}_i(t) \times \frac{\partial \bold{M}_i(t)}{\partial t} + \\
 \frac{J^2_{sd}}{\mu_M} \bigg\{\sum_{p=\mathrm{L,R}}I_{p,i}^\mathrm{D}(t)\bigg\}\bold{M}_i(t) \times \frac{\partial^2 \bold{M}_i(t)}{\partial t^2}\bigg].
\end{multline}
However, the Gilbert damping term prefactor
%=====EQUATION======
\begin{equation} \label{eq:dlambda}
\lambda^\mathrm{D}_i(t) = \frac{J_{sd}^2}{\mu_M}\sum_{p=\mathrm{L,R}}\lambda^\mathrm{D}_{p,i}(t),
\end{equation}
and the magnetic inertia term prefactor
\begin{equation} \label{eq:dinertia}
I^\mathrm{D}_i(t) = \frac{J_{sd}^2}{\mu_M}\sum_{p=\mathrm{L,R}}I^\mathrm{D}_{p,i}(t), 
\end{equation}
in Eq.~\eqref{eq:llg_nu} are now time- and position-dependent. This is in sharp contrast to conventional LLG Eq.~\eqref{eq:llg} employed in classical micromagnetics where Gilbert damping and magnetic inertia prefactors are material specific constants. 

%!!!!!!!!!!!!!!!!!!!!!!!!!!!!!!!!!!!!!!!!!!!!!!!!!!!!!!!!!!!!!!!!!!!!!!!!!!!!!!!!!!!!!!!!!!!!!!!!!!!!!!!!!!!!!!!!!!!!!!!!!!!!!!!!!!!!!!!!!!!!!!!!!!!!!!!!!!!!!!!!!!!!!!!!!!!!!!!
\section{Results and Discussion}\label{sec:results}
\subsection{Single local magnetic moment in an external magnetic field}\label{sec:singlespin}
%!!!!!!!!!!!!!!!!!!!!!!!!!!!!!!!!!!!!!!!!!!!!!!!!!!!!!!!!!!!!!!!!!!!!!!!!!!!!!!!!!!!!!!!!!!!!!!!!!!!!!!!!!!!!!!!!!!!!!!!!!!!!!!!!!!!!!!!!!!!!!!!!!!!!!!!!!!!!!!!!!!!!!!!!!!!!!!!
%=====FIGURE======
\begin{figure}
\includegraphics[scale=0.34]{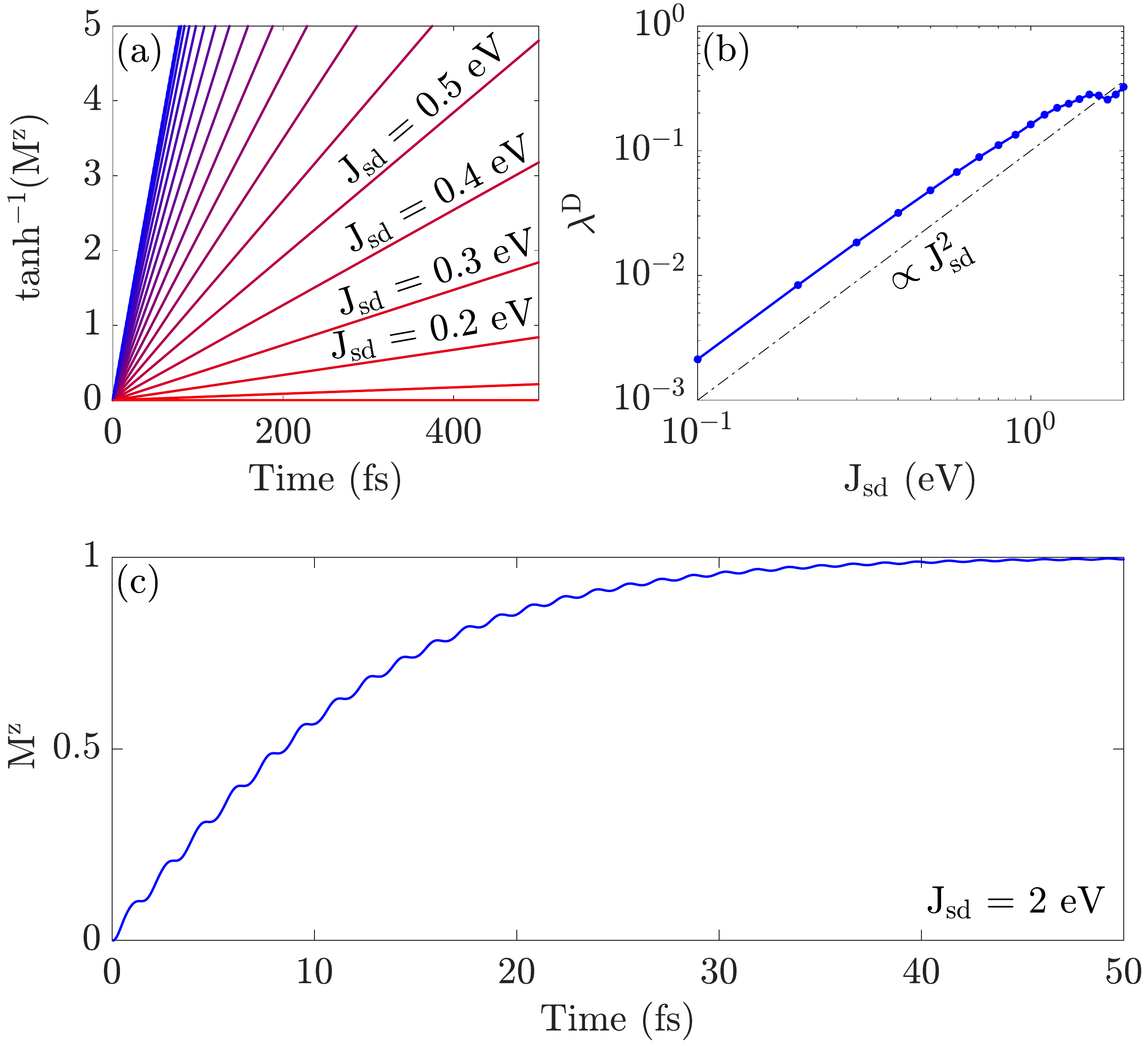}
\caption{ (a) Time dependence of $\tanh^{-1}(M^z)$ for a single local magnetic moment in  Fig.~\ref{fig:fig1}(a) obtained from TDNEGF+LLG simulations. Colors red to blue indicate  increasing $s$-$d$ exchange coupling in steps of \mbox{$0.1$ eV}, ranging from \mbox{$J_{sd} = 0$ eV} to \mbox{$J_{sd} = 1.9$ eV}. (b) The dynamical Gilbert damping parameter in  Eq.~\eqref{eq:dlambda} extracted from panel (a) as a function of $J_{sd}$. (c) Time dependence of $M^z$ component for a single local magnetic moment in  Fig.~\ref{fig:fig1}(a) at large \mbox{$J_{sd} = 2.0$ eV} exhibits nutation as a signature of magnetic inertia. To generate fast magnetization dynamics and reduce simulation time, we use an unrealistically large external magnetic field of strength \mbox{$B = 1000$ T}. The conventional intrinsic Gilbert damping parameter is set to zero, $\lambda_\mathrm{G} = 0$, and the Fermi energy is \mbox{$E_\mathrm{F} = 0$ eV}.}
\label{fig:fig2}
\end{figure}
To compare the dynamics of local magnetic moments in full TDNEGF+LLG quantum-classical simulations vs. conventional LLG classical simulations, we first consider a well-known example~\cite{Evans2014} for which the conventional LLG equation can be analytically solved---a single local magnetic moment which at $t=0$ points along the $+x$-direction and then starts to precesses due to an external magnetic field pointing in the $+z$-direction. Its trajectory is given by~\cite{Evans2014} 
\begin{subequations}\label{eq:sxyz}
%=====EQUATION======
\begin{eqnarray}\label{eq:sx}
M^{x}(t) = \mathrm{sech}  \bigg( \frac{g\lambda_\mathrm{G}B}{1+\lambda_{\mathrm{G}}} t\bigg) \cos \bigg( \frac{gB}{1+\lambda_\mathrm{G}^2} t \bigg),
\end{eqnarray}
%=====EQUATION======
\begin{eqnarray}\label{eq:sy}
M^{y}(t) = \mathrm{sech}  \bigg( \frac{g\lambda_\mathrm{G}B}{1+\lambda_{\mathrm{G}}} t\bigg) \sin \bigg( \frac{gB}{1+\lambda_\mathrm{G}^2} t \bigg),
\end{eqnarray}
%=====EQUATION======
\begin{eqnarray}\label{eq:sz}
M^{z}(t) = \tanh \bigg( \frac{g\lambda_\mathrm{G}B}{1+\lambda_{\mathrm{G}}} t \bigg),
\end{eqnarray}
\end{subequations}
where $\mathbf{B} = (0,0,B)$ is the applied external magnetic field. Thus, if the conventional intrinsic Gilbert damping parameter is set to zero, $\lambda_\mathrm{G} = 0$, then the local magnetic moment precesses steadily around the $z$-axis with \mbox{$M^z\equiv 0$}. On the other hand, for nonzero $\lambda_\mathrm{G} > 0$, the local magnetic moment will relax towards  the direction of magnetic field, i.e., \mbox{$\lim\limits_{t \to \infty}\left(M^{x}(t),M^{y}(t),M^{z}(t) \right)=(0,0,1)$}. Thus, such damped dynamics is signified by a linear $\mathrm{tanh^{-1}}(M^{z})$ vs. time dependence. Figure~\ref{fig:fig2} plots results of TDNEGF+LLG simulations for the same problem. Even though we set conventional intrinsic Gilbert damping to zero, \mbox{$\lambda_{\mathrm{G}}$ = 0}, Fig.~\ref{fig:fig2}(a) shows  linear $\tanh^{-1}(M^z)$ vs. time, independently of the strength of $s$-$d$ exchange coupling as long as \mbox{$J_{sd}\lesssim 2$ eV}. This means that the local magnetic moment is experiencing (time-independent) dynamical Gilbert damping $\lambda^\mathrm{D}\propto J_{sd}^2$,  in accord with Eq.~\eqref{eq:dlambda} and as shown in Fig.~\ref{fig:fig2}(b), which is generated solely by the TDNEGF part of the self-consistent loop within the full TDNEGF+LLG scheme. 

For \mbox{$J_{sd}\gtrsim 2$ eV}, the dynamics of the local magnetic moment also exhibits nutation~\cite{Sayad2015}, as shown in Fig.~\ref{fig:fig2}(c), which is the signature of the magnetic inertia~\cite{Faehnle2011,Bhattacharjee2012,Kikuchi2015,Sayad2016a,Mondal2017,Mondal2018} term $\propto \bold{M}_i\times \partial^2\bold{M}_i/\partial t^2 $ in Eq.~\eqref{eq:llg_nu}. Thus, nutation becomes conspicuous when the dynamics of the local magnetic moments is sufficiently fast, so that $\partial^2\bold{M}_i/\partial t^2 $ is large, as well as when the interaction between the itinerant  and localized spins is sufficiently large.

%Further, we observe that for strong $s$-$d$ exchange coupling (\mbox{$J_{sd} \sim 2$ eV}) the magnetic moment exhibits nutation. In TDNEGF+LLG framework the origin of nutation can be traced back to magnetic inertia term ($\propto \bold{S}_i\times \partial_t^2\bold{S}_i$) in Eq.~\eqref{eq:llg_nu}, whose strength increases as $s$-$d$ exchange coupling increases, which is also predicted by our analytical results in Eq.~\eqref{eq:dinertia}.

%!!!!!!!!!!!!!!!!!!!!!!!!!!!!!!!!!!!!!!!!!!!!!!!!!!!!!!!!!!!!!!!!!!!!!!!!!!!!!!!!!!!!!!!!!!!!!!!!!!!!!!!!!!!!!!!!!!!!!!!!!!!!!!!!!!!!!!!!!!!!!!!!!!!!!!!!!!!!!!!!!!!!!!!!!!!!
\subsection{Multiple exchange-uncoupled local magnetic moments in an external magnetic field}
\label{sec:multiplespin}
%!!!!!!!!!!!!!!!!!!!!!!!!!!!!!!!!!!!!!!!!!!!!!!!!!!!!!!!!!!!!!!!!!!!!!!!!!!!!!!!!!!!!!!!!!!!!!!!!!!!!!!!!!!!!!!!!!!!!!!!!!!!!!!!!!!!!!!!!!!!!!!!!!!!!!!!!!!!!!!!!!!!!!!!!!!!!!
%=====FIGURE======
\begin{figure}
\includegraphics[scale=0.34]{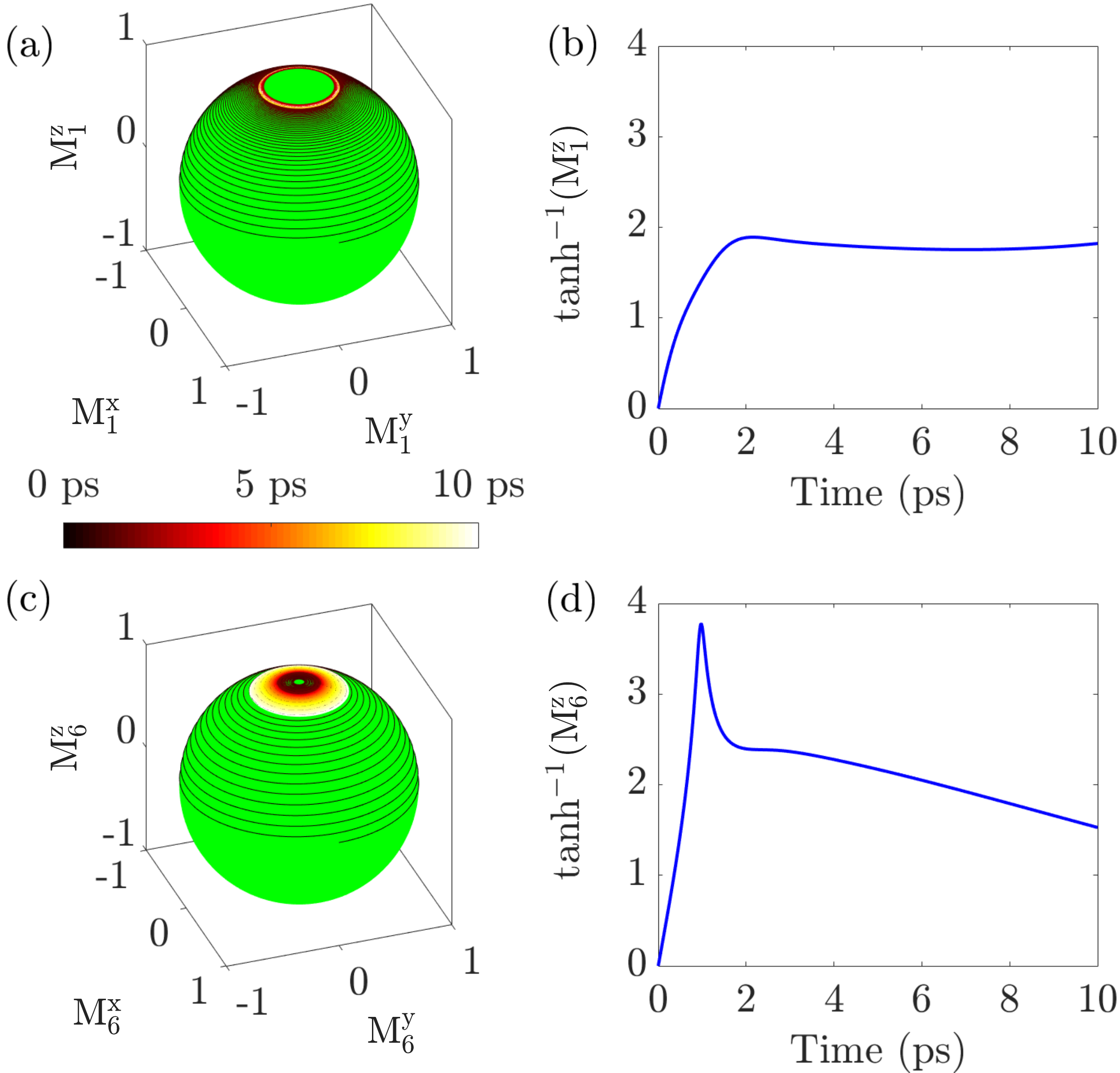}
\caption{TDNEGF+LLG-computed trajectories $(M^{x}(t),M^{y}(t),M^{z}(t))$ on the Bloch sphere of local magnetic moment in the setup of Fig.~\ref{fig:fig1}(b) at: (a) site 1; and (c) site 6.  The total number of local magnetic moments is $N=11$, and they do not interact with each other via exchange coupling [i.e., \mbox{$J=0$ eV} in Eq.~\eqref{eq:classmm}]. Panels (b) and (d) show the corresponding time dependence of $M^z$ component from panels (a) and (c), respectively. The external magnetic field is \mbox{$B=1000$ T}, and the $s$-$d$ exchange coupling strength \mbox{$J_{sd}$ = 0.1 eV} is \textit{nonperturbative} in this setup, therefore, {\em not} allowing us to extract explicitly the dynamical Gilbert damping parameter from Eq.~\eqref{eq:dlambda}. The conventional intrinsic Gilbert damping parameter is  set to zero, $\lambda_\mathrm{G} = 0$,  and the Fermi energy is \mbox{$E_\mathrm{F} = 0$ eV}.}
\label{fig:fig3} 
\end{figure}

In order to examine possible spatial dependence of the dynamical Gilbert damping parameter or emergence of dynamical exchange coupling~\cite{Fransson2010,Saygun2016} between local magnetic moments, we consider a chain of $N=11$ magnetic moments which do not interact with each other ($J=0$) but interact with conduction electron spin (\mbox{$J_{sd} \neq 0$}), as illustrated in Fig.~\ref{fig:fig1}(b). At $t=0$, all magnetic moments point in the $+x$-direction while the external magnetic field is in the $+z$ direction, and the conventional intrinsic Gilbert damping is set to zero, $\lambda_\mathrm{G} = 0$.

Figures~\ref{fig:fig3}(a) and ~\ref{fig:fig3}(c) show the trajectory of selected local magnetic moments ($i=1$ and $6$) on the Bloch sphere for \mbox{$J_{sd} = 0.1$ eV}. In contrast to single local magnetic moment in Fig.~\ref{fig:fig2}(a), for which $\mathrm{tanh^{-1}}(M^z)$ vs. time  is linear using \mbox{$J_{sd} = 0.1$ eV}, we find that in case of multiple exchange-uncoupled magnetic moments this is no longer the case, as demonstrated by Figs.~\ref{fig:fig3}(b) and ~\ref{fig:fig3}(d). Hence, the trajectory followed by these local magnetic moments cannot be described by  Eq.~\eqref{eq:sxyz} so that the conventional-like Gilbert damping parameter cannot be extracted anymore. Thus, such a nonstandard damping of the dynamics of local magnetic moments originates from time-dependence of the dynamical damping parameter $\lambda_i^\mathrm{D}$ in Eq.~\eqref{eq:dlambda}.

\begin{figure}
\includegraphics[scale=0.34]{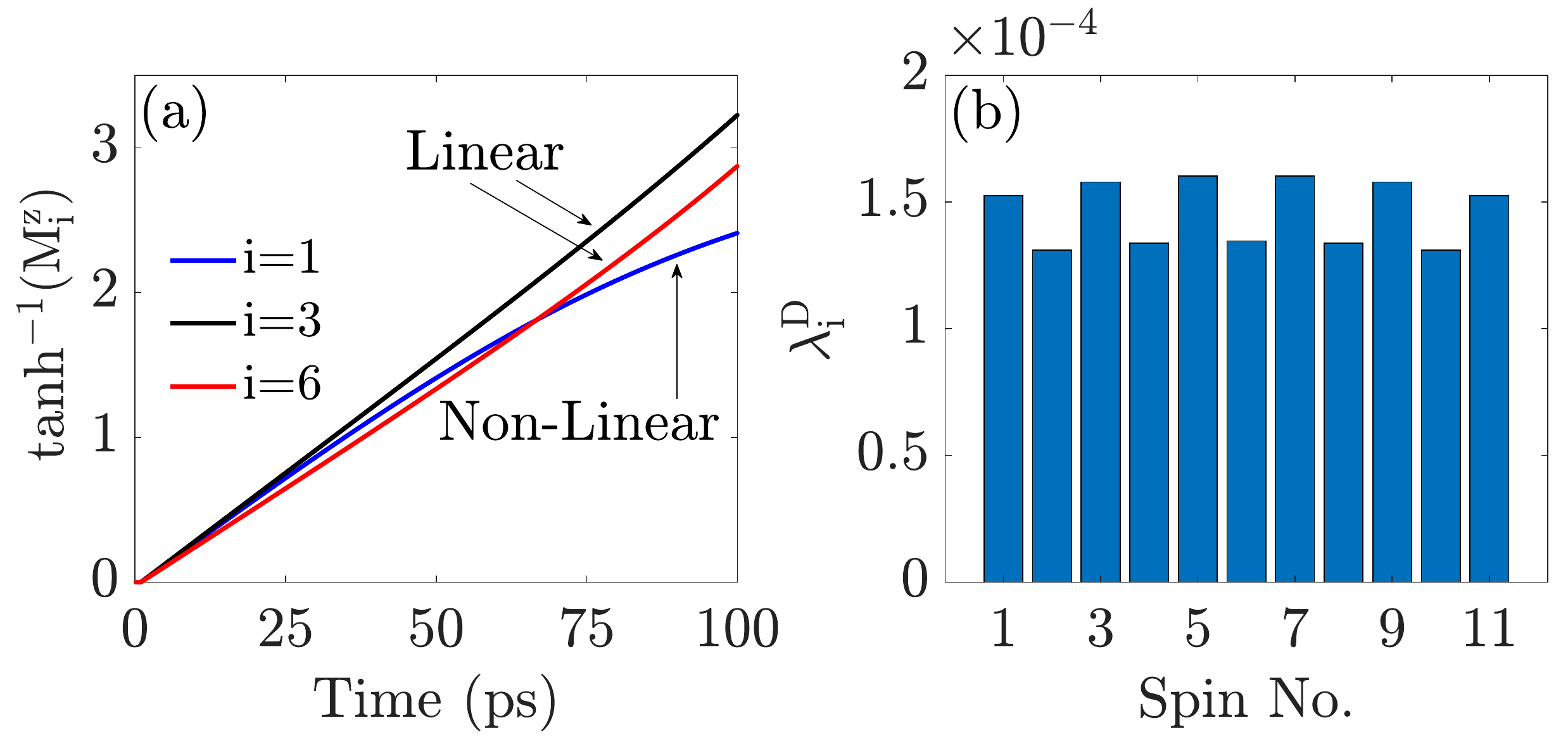}
\caption{(a) TDNEGF+LLG-computed time dependence of $M^{z}$ component of local magnetic moment on sites 1, 3 and 6 in the setup of Fig.~\ref{fig:fig1}(b) with a total of $N=11$ moments. (b) Position dependence  of the dynamical Gilbert damping parameter in Eq.~\eqref{eq:dlambda}. The external magnetic field is \mbox{$B=1000$ T}, and the $s$-$d$ exchange coupling strength \mbox{$J_{sd}$ = 0.01 eV} is \textit{perturbative} in this setup, therefore, allowing us to extract the dynamical Gilbert damping explicitly from Eq.~\eqref{eq:dlambda}. The conventional intrinsic Gilbert damping parameter is  set to zero, $\lambda_\mathrm{G} = 0$, and the Fermi energy is \mbox{$E_\mathrm{F} = 0$ eV}.}
\label{fig:fig4}
\end{figure}

Figure~\ref{fig:fig4}(a) shows $\mathrm{tanh^{-1}}(M^z)$ vs. time for selected local magnetic moments ($i=1,3$ and $6$) and smaller \mbox{$J_{sd}=0.01$ eV}. Although all local magnetic moments follow linear $\mathrm{tanh^{-1}}(M^z)$ vs. time, as predicted by the solution in Eq.~\eqref{eq:sz} of the conventional LLG equation, the dynamical Gilbert damping extracted from Eq.~\eqref{eq:sxyz} changes from site to site as shown in Fig.~\ref{fig:fig4}(b). Furthermore, the linear $\mathrm{tanh^{-1}}(M^z)$ vs. time relation breaks down for times \mbox{$t\gtrsim 50$ ps} at specific sites, which then prevents extracting time-independent $\lambda_i^\mathrm{D}$ at those sites.

%=====FIGURE======

%!!!!!!!!!!!!!!!!!!!!!!!!!!!!!!!!!!!!!!!!!!!!!!!!!!!!!!!!!!!!!!!!!!!!!!!!!!!!!!!!!!!!!!!!!!!!!!!!!!!!!!!!!!!!!!!!!!!!!!!!!!!!!!!!!!!!!!!!!!!!!!!!!!!!!!!!!!!!!!!!!!!!!!!!!!!!!!!
\subsection{Magnetic field-driven motion of a domain wall composed of multiple exchange-coupled local magnetic moments}\label{sec:magneticdw}
%!!!!!!!!!!!!!!!!!!!!!!!!!!!!!!!!!!!!!!!!!!!!!!!!!!!!!!!!!!!!!!!!!!!!!!!!!!!!!!!!!!!!!!!!!!!!!!!!!!!!!!!!!!!!!!!!!!!!!!!!!!!!!!!!!!!!!!!!!!!!!!!!!!!!!!!!!!!!!!!!!!!!!!!!!!!!!!!

%=====FIGURE======
\begin{figure*}
\includegraphics[scale=0.32]{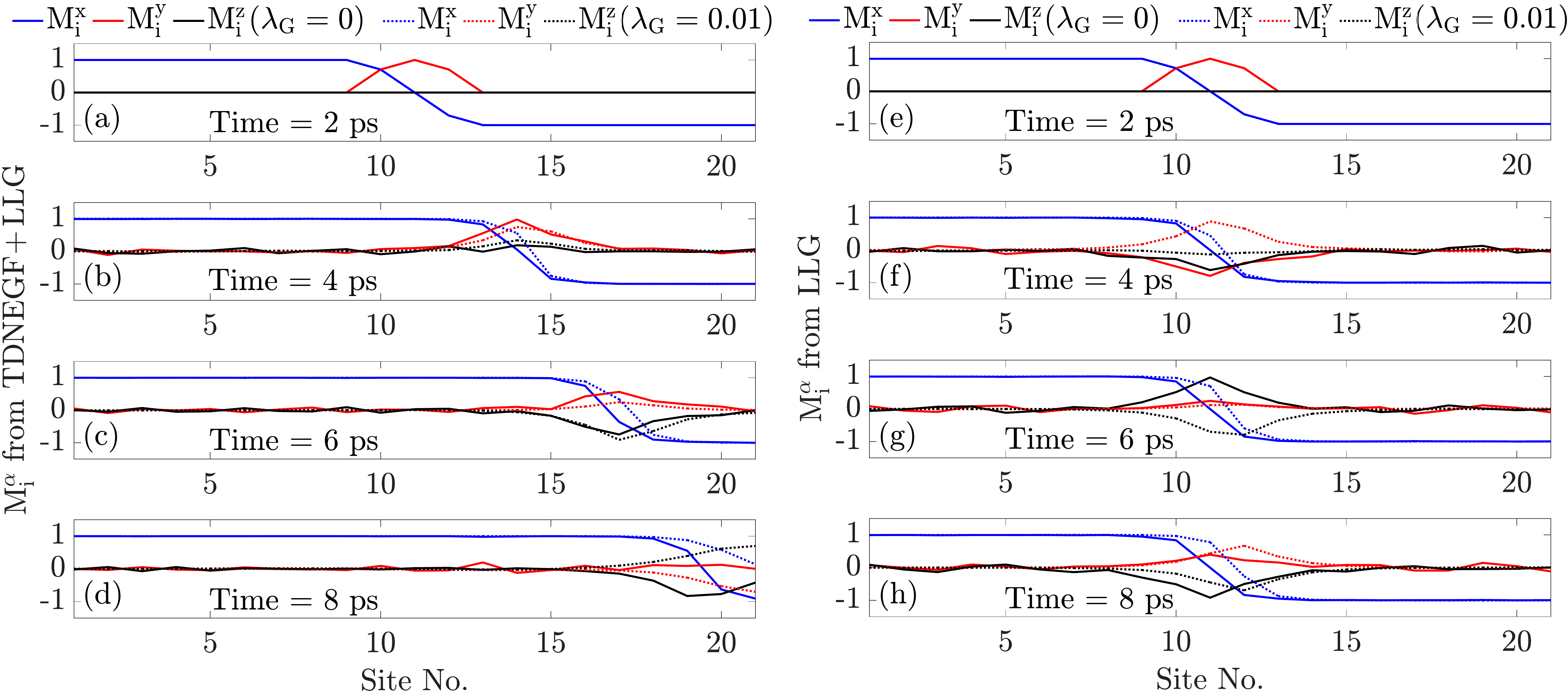}
\caption{(a)--(d) TDNEGF+LLG-computed snapshots of head-to-head DW in the setup of Fig.~\ref{fig:fig1}(c) driven by an external magnetic field of strength \mbox{$B=100$ T} pointing in the $+x$-direction, in the absence ($\lambda_\mathrm{G} = 0$) or presence ($\lambda_\mathrm{G} = 0.01$) of the conventional intrinsic Gilbert damping. Panels (e)--(h) show the corresponding  snapshots computed solely by the conventional LLG Eq.~\eqref{eq:llg} where in the absence ($\lambda_\mathrm{G} = 0$) of the conventional intrinsic Gilbert damping the DW does not move at all. The Heisenberg exchange coupling between local magnetic moments is \mbox{$J=0.01$ eV}; $s$-$d$ exchange coupling between electrons and local magnetic moments is \mbox{$J_{sd}= 0.1$ eV}; magnetic anisotropy (in the $x$-direction) is \mbox{$K=0.01$ eV}; and the Fermi energy of electrons is \mbox{$E_F=-1.9$ eV}. The magnetic field is applied at \mbox{$t=2$ ps}, while prior to that we evolve the conduction electron subsystem with TDNEGF until it reaches the thermodynamic equilibrium where all transient spin and charge currents have decayed to zero.}
\label{fig:fig5}
\end{figure*}
In order to examine difference in predicted dynamics of exchange-coupled local magnetic moments by TDNEGF+LLG framework vs. conventional LLG equation, we consider the simplest example of 1D head-to-head magnetic DW depicted in Fig.~\ref{fig:fig1}(c). Its motion is driven by applying an external magnetic field in the $+x$-direction. Some type of damping mechanism is crucial for the DW to move, as demonstrated by solid lines in Fig.~\ref{fig:fig5}(e)--(h), obtained by solving the conventional LLG equation with $\lambda_\mathrm{G} = 0$, which show how local magnetic moments precess around the magnetic field but without net displacement of the center of the DW.

On the other hand, even though we set $\lambda_\mathrm{G} = 0$ in TDNEGF+LLG simulations in Fig.~\ref{fig:fig5}(a)--(d), the center of the DW moves to the right due to dynamically generated time-retarded damping encoded by the memory kernel in Eq.~\eqref{eq:llg_mem}. Including the conventional intrinsic Gilbert damping, $\lambda_\mathrm{G} = 0.01$ as often used in micromagnetic simulations of DW along magnetic nanowires~\cite{Thiaville2005,Thiaville2007,Taniguchi2015}, changes only slightly the result of TDNEGF+LLG simulations which demonstrates that the effective dynamical Gilbert damping (which is also time-dependent) is about an order of magnitude larger than $\lambda_\mathrm{G}$. This is also reflected in the DW velocity being much larger in TDNEGF+LLG simulations with $\lambda_\mathrm{G}=0$ in Fig.~\ref{fig:fig5}(a)--(d) than in the conventional LLG equation simulations with $\lambda_\mathrm{G}=0.01$ in Fig.~\ref{fig:fig5}(e)--(h).

It has been predicted theoretically~\cite{Berger1986,Volovik1987,Stern1992,Barnes2007,Zhang2009b,Duine2008,Tserkovnyak2008a} and confirmed experimentally~\cite{Yang2009} that a moving DW will pump charge current even in the absence of any applied bias voltage. The corresponding open circuit pumping voltage in the so-called spin motive force (SMF) theory~\cite{Barnes2007,Zhang2009b} is given by 
\begin{subequations}\label{eq:smf_eqn}
\begin{equation}\label{eq:vsmf}
V_{\mathrm{SMF}} = \frac{1}{G_0}\int j_{x}dx,
\end{equation}
\begin{equation}\label{eq:smfcharge}
j_\alpha (\mathbf{r}) = \frac{P \sigma_0 \hbar}{2e} [\partial_t \mathbf{m}(\mathbf{r},t) \times \partial_\alpha \mathbf{m}(\mathbf{r},t)] \cdot \mathbf{m}(\mathbf{r},t),
\end{equation}
\end{subequations}
where $j_x$ is the pumped local charge current along the $x$-axis. Here $\sigma_0=\sigma^\uparrow + \sigma^\downarrow$ is the total conductivity; \mbox{$P=(\sigma^\uparrow - \sigma^\downarrow)/(\sigma^\uparrow + \sigma^\downarrow)$} is the spin polarization of the ferromagnet; and $\partial_t = \partial/\partial t$.
Equation~\eqref{eq:smf_eqn} is typically combined~\cite{Ohe2009,Shimada2015,Yamane2016} with classical micromagnetics which supplies $\bold{M}_i(t)$ that is then plugged into the discretized version~\cite{Petrovic2018} %
\begin{eqnarray}\label{eq:smfdiscrete}
j_x(i) & \propto & \frac{1}{a} [\partial_t \mathbf{M}_i(t) \times (\mathbf{M}_{i+1}(t)-\mathbf{M}_i(t))] \cdot \mathbf{M}_i(t)  \nonumber \\ 
 & \propto & \frac{1}{a}[\partial \mathbf{M}_i(t) \times \mathbf{M}_{i+1}(t)]\cdot \mathbf{M}_i(t).
\end{eqnarray} 
of Eq.~\eqref{eq:smfcharge}. We denote this approach as SMF $\leftarrow$ LLG, which is perturbative in nature~\cite{Duine2008,Freimuth2018} since it considers only the lowest temporal and spatial derivatives.  

On the other hand, the same pumping voltage can be computed nonperturbatively 
\begin{equation}\label{eq:vtdnegf}
V_{\mathrm{TDNEGF}} = \frac{I_p(t)}{G(t)},
\end{equation}
using TDNEGF expression for charge current in lead $p$ in Eq.~\ref{eq:c_curr}, where TDNEGF calculations are coupled to LLG calculations either self-consistently (i.e., by using TDNEGF $\leftrightarrows$ LLG) or non-self-consistently (i.e., by using TDNEGF $\leftarrow$ LLG). Here, $G(t)$ is the conductance computed using the Landauer formula applied to two-terminal devices with a frozen at time $t$ texture of local magnetic moments.
%=====FIGURE======
\begin{figure}
\hspace{-0.2cm}\includegraphics[scale=0.32]{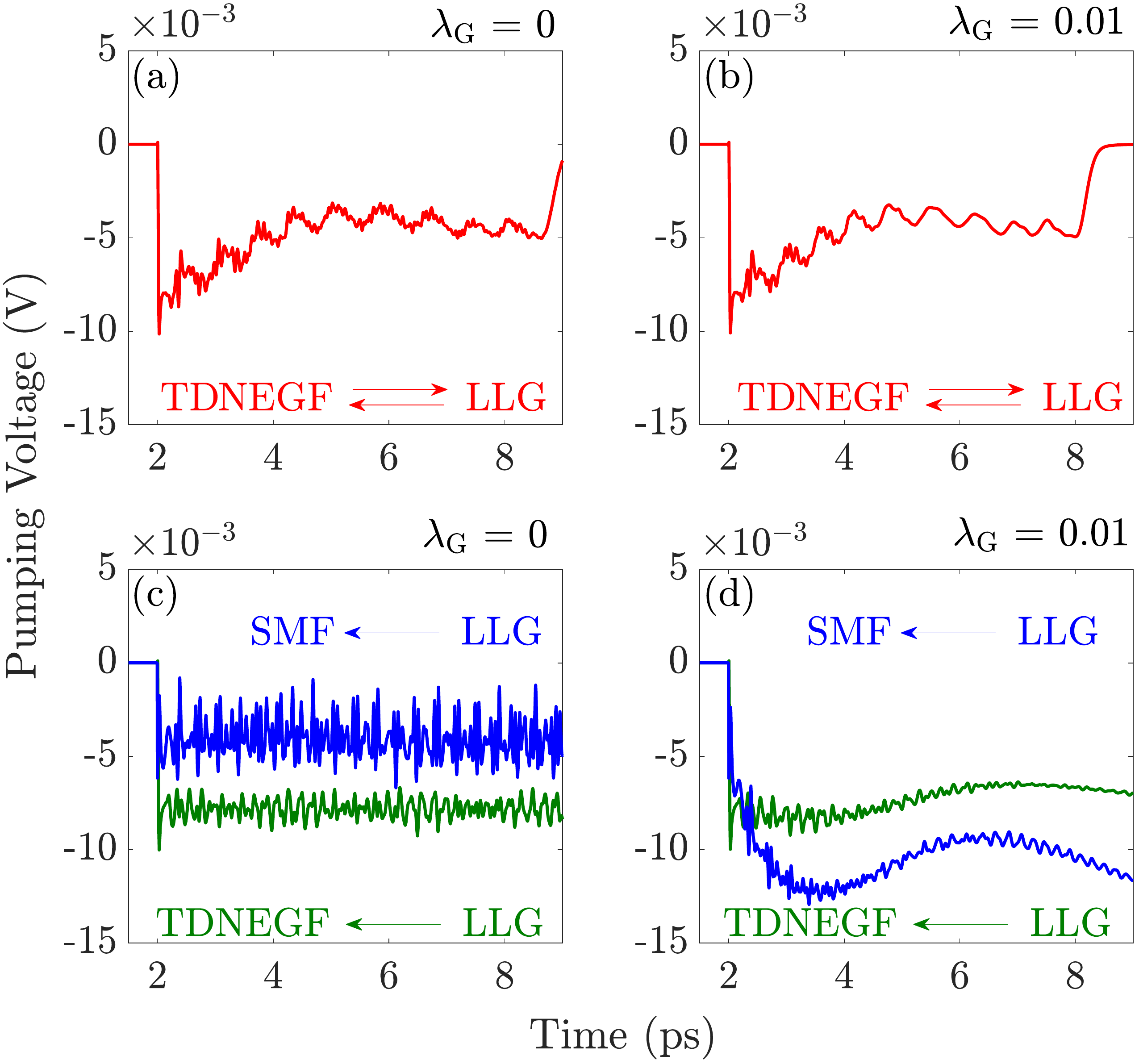}
\caption{Time dependence of pumping voltage generated by the DW motion depicted in Fig.~\ref{fig:fig5}(a)--(d) for: (a) $\lambda_\mathrm{G} = 0$; (b) $\lambda_\mathrm{G} = 0.01$. In panels (a) and (b) local magnetic moments  evolve in time by the full TDNEGF+LLG framework where the arrows indicate how TDNEGF sends nonequilibrium electronic spin density into the LLG equation which, in turn, sends trajectories of local magnetic moments into TDNEGF. Time dependence of pumping voltage generated by DW motion depicted in  Fig.~\ref{fig:fig5}(e)--(h) for: (c) $\lambda_\mathrm{G} = 0$; (d) $\lambda_\mathrm{G} = 0.01$. In  panels (c) and (d) local magnetic moments  evolve in time using the conventional LLG equation which sends their trajectories into either TDNEGF (green) or SMF formulas (blue) in Eq.~\eqref{eq:vtdnegf} or Eq.~\eqref{eq:smf_eqn}, respectively, to obtain the corresponding pumping voltage.}
\label{fig:fig6}
\end{figure}

Figures~\ref{fig:fig6}(a) and ~\ref{fig:fig6}(b) plot the pumping voltage calculated by TDNEGF $\leftrightarrows$ LLG  for DW motion shown in Fig.~\ref{fig:fig5}(a)--(d) in the absence or presence of conventional Gilbert damping, respectively. The two cases are virtually identical due to an order of magnitude larger dynamical Gilbert damping that is automatically generated by TDNEGF $\leftrightarrows$ LLG in both Figs.~\ref{fig:fig6}(a) and ~\ref{fig:fig6}(b). The nonperturbative results in Figs.~\ref{fig:fig6}(a) and Fig.~\ref{fig:fig6}(b) are quite different from SMF $\leftarrow$ LLG predictions in Figs.~\ref{fig:fig6}(c) and Fig.~\ref{fig:fig6}(d), respectively. This is due to both failure of Eqs.~\eqref{eq:smf_eqn} and~\eqref{eq:smfdiscrete} to describe noncoplanar and noncollinear magnetic textures with neighboring local magnetic moments tilted by more than $10^\circ$~\cite{Petrovic2018} and lack of dynamical Gilbert damping in SMF $\leftarrow$ LLG simulations~\cite{Ohe2009,Shimada2015,Yamane2016}. The latter effect is also emphasized by the inability of TDNEGF $\leftarrow$ LLG in  Figs.~\ref{fig:fig6}(c) and Fig.~\ref{fig:fig6}(d) to reproduce the results of self-consistent TDNEGF $\leftrightarrows$ LLG in Figs.~\ref{fig:fig6}(a) and Fig.~\ref{fig:fig6}(b), respectively.

%!!!!!!!!!!!!!!!!!!!!!!!!!!!!!!!!!!!!!!!!!!!!!!!!!!!!!!!!!!!!!!!!!!!!!!!!!!!!!!!!!!!!!!!!!!!!!!!!!!!!!!!!!!!!!!!!!!!!!!!!!!!!!!!!!!!!!!!!!!!!!!!!!!!!!!!!!!!!!!!!!!!!!!!!!!!!!!!
\section{Conclusions}\label{sec:conclusions}
%!!!!!!!!!!!!!!!!!!!!!!!!!!!!!!!!!!!!!!!!!!!!!!!!!!!!!!!!!!!!!!!!!!!!!!!!!!!!!!!!!!!!!!!!!!!!!!!!!!!!!!!!!!!!!!!!!!!!!!!!!!!!!!!!!!!!!!!!!!!!!!!!!!!!!!!!!!!!!!!!!!!!!!!!!!!!!!!
In conclusion, we delineated a hierarchy of theoretical descriptions of a nonequilibrium quantum many-body system in which conduction electron spins interact with local magnetic moments  within a ferromagnetic layer sandwiched between normal metal electrodes. On the top of the hierarchy is a fully quantum approach, for both electrons and local magnetic moments, whose computational complexity (using either original spin operators~\cite{Sayad2016,Mondal2018a} for local magnetic moments, or their mapping to bosonic operators in order to enable application of many-body perturbation theory within the NEGF formalism~\cite{Mahfouzi2014}) makes it impractical for systems containing large number of local magnetic moments. The next approach in the hierarchy is computationally much less expensive quantum-classical hybrid~\cite{Elze2012a}  based on self-consistent coupling~\cite{Petrovic2018} of TDNEGF (which can be implemented using algorithms that scale linearly  with both system size and simulation time~\cite{Gaury2014,Weston2016,Popescu2016}) with classical LLG equation for local magnetic moments. Such TDNEGF+LLG approach is numerically exact and, therefore, nonperturbative in the strength of electron-spin/local-magnetic-moment interaction, speed of local magnetic moment dynamics and degree of noncollinearity between them. Even  though electron dynamics is much faster than localized spin dynamics, the most general situation cannot be handled by integrating out~\cite{Onoda2006d,Nunez2008} the conduction electron degrees of freedom and by focusing only on the LLG-type equation where a much larger time step can be used to propagate spins only. 

Nevertheless, in the limit~\cite{Onoda2006d,Zhang2004a} of weak electron-spin/local-magnetic-moment interaction [i.e., small $J_{sd}$ in Eqs.~\eqref{eq:ham} and~\eqref{eq:classmm}] one can derive analytically a type of generalized LLG equation~\cite{Onoda2006d,Sayad2015,Hammar2016,Hammar2017} for each local magnetic moment which is next approach in the hierarchy that sheds light onto different effects included in the numerically exact TDNEGF+LLG scheme. Instead of the conventional Gilbert damping term in Eq.~\eqref{eq:llg}, the generalized LLG equation we derive as Eq.~\eqref{eq:llg_mem} contains a microscopically determined memory kernel which describes time-retardation effects generated by the coupling to TDNEGF. Fundamentally, the memory kernel is due to the fact that electron spin can never follow instantaneously change in the orientation of the local magnetic moments~\cite{Sayad2015}. In the limit of slow dynamics of local magnetic moments, one can further expand the memory kernel into a Taylor series to obtain the final approach within the hierarchy whose LLG Eq.~\eqref{eq:llg_nu} is akin to the conventional one, but which contains both Gilbert damping (proportional to first time derivative of local magnetization) and magnetic inertia terms (proportional to second time derivative of local magnetization) with {\em time-dependent} parameters instead of usually assumed materials specific constants. 

Using three simple examples---single or multiple local magnetic moments precessing in an external magnetic field or magnetic-field-driven magnetic DW motion---we demonstrate the importance of dynamically induced damping which operates even if conventional static Gilbert damping is set to zero. In the case of field-driven magnetic DW motion, we can estimate that the  strength of dynamical damping is effectively {\em an order of magnitude larger} than typically assumed~\cite{Thiaville2005,Thiaville2007,Taniguchi2015} conventional static Gilbert damping $\lambda_\mathrm{G} \simeq 0.01$ in classical micromagnetic simulations of magnetic nanowires. In addition, we show that charge pumping by the dynamics of noncoplanar and noncollinear magnetic textures, which is outside of the scope of pure micromagnetic simulations but it is often described by combining~\cite{Ohe2009,Shimada2015,Yamane2016} them with the SMF theory formula~\cite{Barnes2007,Zhang2009b}, requires to take into account both the dynamical Gilbert damping and possibly large angle between neighboring local magnetic moments in order to reproduce numerically exact results of TDNEGF+LLG scheme.

%======================================ACKNOWLEDGEMENTS==============================
\begin{acknowledgments}
This work was supported by NSF Grant No. ECCS 150909.
\end{acknowledgments}

%==========================================APPENDIX==================================
\appendix
\numberwithin{equation}{section}
\begin{widetext}
\section{Derivation of Memory Kernel in LLG equation self-consistently coupled to TDNEGF}\label{sec:app}
In this Appendix, we provide a detailed derivation of the memory kernel in Eq.~\eqref{eq:llg_mem}. To obtain the  perturbative equation of motion for local magnetic moments we start from Landau-Lifshitz Eq.~\eqref{eq:llg_sd} where the effective magnetic field can be written as
\begin{equation}\label{eq:B0B1}
\bold{B}^\mathrm{eff}_i(t) = \bold{B}^{\mathrm{eff},0}_i(t) + J_{sd} \langle \hat{\mathbf{s}} \rangle^i(t).
\end{equation}
The nonequilibrium spin density is expanded up to terms linear in $J_{sd}$ using Eq.~\eqref{eq:dm_exp}
%=====EQUATION======
\begin{equation}  \label{eq:sd}
\langle \hat{\mathbf{s}} \rangle^i(t) = \frac{\hbar}{2}\Tr[\DM_\mathrm{neq}(t) \ket{i}\bra{i}\otimes\boldsymbol{\sigma}] - \langle \hat{\mathbf{s}} \rangle_{\mathrm{eq}}^i \approx \frac{\hbar}{2}\Tr\bigg[\{\DM_0(t) + J_{sd}\DM_1(t)\} \ket{i}\bra{i}\otimes\boldsymbol{\sigma}\bigg] - \langle \hat{\mathbf{s}} \rangle_{\mathrm{eq}}^i = J_{sd}\frac{\hbar}{2}\Tr[\DM_1(t) \ket{i}\bra{i}\otimes\boldsymbol{\sigma}] -\langle \hat{\mathbf{s}} \rangle_{\mathrm{eq}}^i.
\end{equation}
Here $\langle \hat{\mathbf{s}} \rangle_{\mathrm{eq}}^i$ is the equilibrium electronic spin density i.e., $\langle \hat{\mathbf{s}} \rangle_{\mathrm{eq}}^i = (\hbar/2)\Tr\, [\DM_\mathrm{eq}\ket{i}\bra{i}\otimes\boldsymbol{\sigma}]$. Furthermore, the electronic spin density in the zeroth order must vanish, i.e., $\Tr\, [\boldsymbol{\rho}_0(t) \ket{i}\bra{i}\otimes\boldsymbol{\sigma}] = 0$ since for $J_{sd}=0$ electrons are not spin-polarized. Hence, we can write Eq.~\eqref{eq:llg_sd} as
\begin{equation} \label{eq:llg_expand}
\frac{\partial \bold{M}_i(t)}{\partial t} = -g\bold{M}_i(t)\times \bigg[\bold{B}^{\mathrm{eff,0}}_i(t) + J_{sd}^2 \frac{\hbar}{2}\Tr[\DM_1(t)\ket{i}\bra{i}\otimes \boldsymbol{\sigma}] - J_{sd}\langle \hat{\mathbf{s}} \rangle_{\mathrm{eq}}^i \bigg].
\end{equation}
To obtain analytical results, we assume that the equilibrium spin density follows the direction of local magnetic moments, so that $\bold{M}_i(t)\times \langle \hat{\mathbf{s}} \rangle_{\mathrm{eq}}^i=0$. By expanding Eq.~\eqref{eq:dme} we obtain
%=====EQUATION======
\begin{equation}  \label{eq:dm_0}
i\hbar\frac{\partial \DM_0(t)}{\partial t} = [\bold{H}_0(t),\DM_0(t)] + \sum_{p = \mathrm{L,R}} i [\bold{\Pi}^{(0)}_{p}(t) + \bold{\Pi}^{(0)\dagger}_{p}(t) ] ,
\end{equation}
%=====EQUATION======
and
\begin{equation} \label{eq:dm_1}
i\hbar\frac{\partial \DM_1(t)}{\partial t} = [\bold{H}_1(t),\DM_0(t)] + i \sum_{p=\mathrm{L,R}} [\bold{\Pi}_{p}^{(1)}(t) + \bold{\Pi }_{p}^{(1)\dagger}(t)],   
\end{equation}
where $\bold{H}_1(t) = -\sum_{i}\ket{i}\bra{i}\otimes \boldsymbol{\sigma}\cdot \bold{M}_i(t)$. One can formally integrate Eq.~\eqref{eq:dm_1} which leads to 
\begin{equation} \label{eq:sd_1}
\frac{\hbar}{2}\Tr[\DM_1(t) \ket{i}\bra{i}\otimes\boldsymbol{\sigma}] = \sum_{p = \mathrm{L,R}}\frac{1}{2} \int \limits_{-\infty}^{t}dt' \Tr\bigg[ \{\bold{\Pi}^{(1)}_{p}(t')+\bold{\Pi}^{(1)\dagger}_{p}(t')\} \ket{i}\bra{i}\otimes\boldsymbol{\sigma}\bigg].
\end{equation}
which requires to find an expression for $\bold{\Pi}_{p}^{(1)}(t')$. Using Eq.~\eqref{eq:currm} and the fact that lead self-energy matrices do not depend on $J_{sd}$ leads to 
%=====EQUATION======
\begin{equation} \label{eq:pi_int}
\bold{\Pi}_{p}^{(1)}(t') =  \int \limits_{-\infty}^{t'} dt_1 [\bold{G}^{>}_1(t',t_1)\bold{\Sigma}^<_{p}(t_1,t') - \bold{G}^{<}_1(t',t_1)\bold{\Sigma}_{p}^>(t_1,t')].
\end{equation}
Equations~\eqref{eq:gfe} and ~\eqref{eq:gf_exp} can be formally integrated to yield lesser and greater GFs in Eq.~\eqref{eq:pi_int}
%=====EQUATION======
\begin{equation}\label{eq:gf_int}
\bold{G}^{>,<}_1(t',t_1) = \frac{1}{i\hbar}\bigg(\int\limits_{-\infty}^{t'}dt''\bold{H}_1(t'')\bold{G}^{>,<}_0(t'',t_1) + \int \limits_{-\infty}^{t'}dt'' \int \limits_{-\infty}^{+\infty}dt_2 \bigg[ \bold{\Sigma}^r_{\mathrm{tot}}(t'',t_2)\bold{G}^{>,<}_1(t_2,t'') + \bold{\Sigma}^{>,<}_{\mathrm{tot}}(t'',t_2)\bold{G}^{a}_1(t_2,t'')\bigg]\bigg).
\end{equation}
We further assume that the active region in Fig.~\ref{fig:fig1} is weakly coupled with semi-infinite leads and, therefore, macroscopic reservoirs into which they terminate. This means that after we substitute Eq.~\eqref{eq:gf_int} into Eq.~\eqref{eq:pi_int} we can keep only those terms that are linear in the self-energy 
%=====EQUATION======
\begin{align}
\bold{\Pi}_{p}^{(1)}(t') &=  \frac{1}{2i}\int \limits_{-\infty}^{t'}dt''\bold{H}_1(t'') \int \limits_{-\infty}^{t'} dt_1\bigg[\bold{G}^{>}_0(t'',t_1)\bold{\Sigma}^{<}_{p}(t_1,t') - \bold{G}^{<}_0(t'',t_1)\bold{\Sigma}_{p}^>(t_1,t') \bigg] \\
&= \frac{i}{2} \sum_{i} \int \limits_{-\infty}^{t'}dt''\ket{i}\bra{i}\otimes \boldsymbol{\sigma}\cdot \bold{M}_i(t'')\int\limits_{-\infty}^{t'} dt_1\bigg[\bold{G}^{>}_0(t'',t_1)\bold{\Sigma}^{<}_{p}(t_1,t') - \bold{G}^{<}_0(t'',t_1)\bold{\Sigma}_{p}^>(t_1,t') \bigg] \\
&=  i \sum_{i} \int\limits_{-\infty}^{t'}dt'' \ket{i}\bra{i}\otimes\boldsymbol{\sigma}\cdot \bold{M}_i(t'') \bold{A}^{0}_{p}(t'',t'),\label{eq:pi_mat}
\end{align}
where $\bold{A}^{0}_{p}(t'',t')$ is an operator constructed out of the zeroth order terms in the expansion of GFs shown in Eq.~\eqref{eq:gf_exp}
%=====EQUATION======
\begin{equation}\label{eq:aop}
\bold{A}^{0}_{p}(t'',t') \equiv  \frac{i}{2}\int\limits_{-\infty}^{t'} dt_1\bigg[\bold{G}^{>}_0(t'',t_1)\bold{\Sigma}^{<}_{p}(t_1,t') - \bold{G}^{<}_0(t'',t_1)\bold{\Sigma}_{p}^>(t_1,t') \bigg].
\end{equation}
By plugging in Eqs.~\eqref{eq:pi_mat} and ~\eqref{eq:aop} into Eq.~\eqref{eq:sd_1} we obtain
%=====EQUATION======
\begin{equation}
\frac{\hbar}{2}\Tr[\DM_1(t)\ket{i}\bra{i}\otimes\hat{\sigma}^{\mu}] = \sum_{p=\mathrm{L,R}}\sum_{j}\sum_{\nu} \int \limits_{-\infty}^tdt' \int \limits_{-\infty}^{t'}dt'' \mathrm{M}^{\nu}_j(t'')\Tr \bigg[ \ket{j}\bra{j}\otimes\hat{\sigma}^{\nu}\{\bold{A}^{0}_{p}(t'',t') + \bold{A}^{0\dagger}_{p}(t'',t') \}\ket{i}\bra{i}\otimes\sigma^{\mu}\bigg].
\end{equation}
Since $\bold{A}^{0}_p(t'',t')$ is an operator constructed from the zeroth order GFs,  it can be written in the followin form
%=====EQUATION======
\begin{equation}
\bold{A}_{p}^{0}(t'',t') = \frac{1}{2}\sum_{mn}\mathrm{A}^{p}_{mn}(t'',t')\ket{m}\bra{n}\otimes\id_2,
\end{equation}
where $\id_2$ is a $2 \times 2$ identity matrix. Using this it is easy to show that
%=====EQUATION======
\begin{equation}\label{eq:rho_1_eval}
\frac{\hbar}{2}\Tr[\DM_1(t)\ket{i}\bra{i}\otimes\boldsymbol{\sigma}] = \sum_{p=\mathrm{L,R}} \int\limits_{-\infty}^{+\infty}\Theta(t-t')dt'\int\limits_{-\infty}^{+\infty}dt''  \bold{M}_i(t'') \{ \mathrm{A}^{
p}_{ii}(t'',t') + \mathrm{A}^{
p*}_{ii}(t'',t')\}\Theta(t'-t''),
\end{equation}
and
%=====EQUATION======
\begin{equation}
\mathrm{K}^{p}_i(t'',t) = \int \limits_{-\infty}^{+\infty}dt' \Theta(t-t')\Theta(t'-t'')\mathrm{A}^{p}_{ii}(t'',t').
\end{equation}
By plugging in   Eq.~\eqref{eq:rho_1_eval}  into the second term on the right hand side of Eq.~\eqref{eq:llg_expand}, we finally obtain Eq.~\eqref{eq:llg_mem} of the main text.
\end{widetext}

%=======================================BIBLIOGRAPHY======================================
%\bibliographystyle{C:/BIBTEX/prsty}
%\bibliography{qttg}

\begin{thebibliography}{10}
	
	\bibitem{Bertotti2009}
	G. Bertotti, I.~D. Mayergoyz, and C. Serpico, {\em Nonlinear magnetization dynamics in nanosystems} (Elsevier, Amsterdam, 2009).
	
	\bibitem{Wieser2013}
	R. Wieser, Phys. Rev. Lett. {\bf 110},  147201  (2013).
	
	\bibitem{Wieser2015}
	R. Wieser, Euro. Phys. J. B {\bf 88},  77  (2015).
	
	\bibitem{Kumar2017}
	D. Kumar and A.~O. Adeyeye, J. Phys. D: Appl. Phys. {\bf 50}, 343001  (2017).
	
	\bibitem{Evans2014}
    R. F. L. Evans, W. J. Fan, P. Chureemart, T. A. Ostler, M. O. A. Ellis, and R. W. Chantrell, J. Phys.: Condens. Matter {\bf 26}, 103202  (2014).
	
	\bibitem{Nunez2008}
	A.~S. N\'u\~nez and R.~A. Duine, Phys. Rev. B {\bf 77},  054401  (2008).
	
	\bibitem{Kambersky2007}
	V. Kambersk\'y, Phys. Rev. B {\bf 76},  134416  (2007).
	
	\bibitem{Gilmore2007}
	K. Gilmore, Y.~U. Idzerda, and M.~D. Stiles, Phys. Rev. Lett. {\bf 99},  027204 (2007).
	
	\bibitem{Mahfouzi2017a}
	F. Mahfouzi, J. Kim, and N. Kioussis, Phys. Rev. B {\bf 96},  214421  (2017).
	
	\bibitem{Ralph2008}
	D. Ralph and M. Stiles, J. Magn. Magn. Mater. {\bf 320}, 1190  (2008).
	
	\bibitem{Tserkovnyak2005}
	Y. Tserkovnyak, A. Brataas, G.~E.~W. Bauer, and B.~I. Halperin, Rev. Mod. Phys. {\bf 77},  1375  (2005).
	
	\bibitem{Zhang2009b}
	S. Zhang and S.~S.-L. Zhang, Phys. Rev. Lett. {\bf 102},  086601  (2009).
	
	\bibitem{Wong2009}
	C.~H. Wong and Y. Tserkovnyak, Phys. Rev. B {\bf 80},  184411  (2009).
	
	\bibitem{Tserkovnyak2009}
	Y. Tserkovnyak, E.~M. Hankiewicz, and G. Vignale, Phys. Rev. B {\bf 79}, 094415  (2009).
	
	\bibitem{Kim2012b}
	K.-W. Kim, J.-H. Moon, K.-J. Lee, and H.-W. Lee, Phys. Rev. Lett. {\bf 108}, 217202  (2012).
	
	\bibitem{Nembach2013}
	H.~T. Nembach, J.~M. Shaw, C.~T. Boone, and T.~J. Silva, Phys. Rev. Lett. {\bf 110},  117201  (2013).
	
	\bibitem{Weindler2014}
	T. Weindler, H. G. Bauer, R. Islinger, B. Boehm, J.-Y. Chauleau, and C. H. Back, Phys. Rev. Lett. {\bf 113},  237204  (2014).
	
	\bibitem{Li2016a}
	Y. Li and W.~E. Bailey, Phys. Rev. Lett. {\bf 116},  117602  (2016).
	
	\bibitem{Faehnle2011}
	M. F\"ahnle, D. Steiauf, and C. Illg, Phys. Rev. B {\bf 84},  172403  (2011).
	
	\bibitem{Bhattacharjee2012}
	S. Bhattacharjee, L. Nordstr\"om, and J. Fransson, Phys. Rev. Lett. {\bf 108}, 057204  (2012).
	
	\bibitem{Kikuchi2015}
	T. Kikuchi and G. Tatara, Phys. Rev. B {\bf 92},  184410  (2015).
	
	\bibitem{Sayad2016a}
	M. Sayad, R. Rausch, and M. Potthoff, EPL (Europhysics Letters) {\bf 116}, 17001  (2016).
	
	\bibitem{Mondal2017}
	R. Mondal, M. Berritta, A.~K. Nandy, and P.~M. Oppeneer, Phys. Rev. B {\bf 96}, 024425  (2017).
	
	\bibitem{Mondal2018}
	R. Mondal, M. Berritta, and P.~M. Oppeneer, J. Phys.: Condens. Matter {\bf 30},  265801  (2018).
	
	\bibitem{Li2015b}
	Y. Li, A.-L. Barra, S. Auffret, U. Ebels, and W. E. Bailey, Phys. Rev. B {\bf 92},  140413  (2015).
	
	\bibitem{Carva2007}
	K. Carva and I. Turek, Phys. Rev. B {\bf 76},  104409  (2007).
	
	\bibitem{Liu2014a}
	Y. Liu, Z. Yuan, R. J. H. Wesselink, A. A. Starikov, and P. J. Kelly, Phys. Rev. Lett. {\bf 113},  207202  (2014).
	
	\bibitem{Wang2008b}
	S. Wang, Y. Xu, and K. Xia, Phys. Rev. B {\bf 77},  184430  (2008).
	
	\bibitem{Ellis2017}
	M.~O.~A. Ellis, M. Stamenova, and S. Sanvito, Phys. Rev. B {\bf 96},  224410
	(2017).
	
	\bibitem{Nikolic2018}
	B.~K. Nikoli\'{c}, K. Dolui, M. D. Petrovi\'{c}, P. Plech\'{a}\v{c}, T. Markussen, and K. Stokbro, in W. Andreoni and S. Yip (eds.) {\em Handbook of Materials Modeling} (Springer, Cham 2018); {\tt arXiv.org/abs/1801.05793}.
	
	\bibitem{Thonig2017}
	D. Thonig, O. Eriksson, and M. Pereiro, Sci. Rep. {\bf 7},  931 (2017).
	
	\bibitem{Bose2011}
	T. Bose and S. Trimper, Phys. Rev. B {\bf 83},  134434  (2011).
	
	\bibitem{Thonig2015}
	D. Thonig, J. Henk, and O. Eriksson, Phys. Rev. B {\bf 92},  104403  (2015).
	
	\bibitem{Onoda2006d}
	M. Onoda and N. Nagaosa, Phys. Rev. Lett. {\bf 96},  066603  (2006).
	
	\bibitem{Sayad2015}
	M. Sayad and M. Potthoff, New J. Phys. {\bf 17},  113058  (2015).
	
	\bibitem{Hammar2016}
	H. Hammar and J. Fransson, Phys. Rev. B {\bf 94},  054311  (2016).
	
	\bibitem{Hammar2017}
	H. Hammar and J. Fransson, Phys. Rev. B {\bf 96},  214401  (2017).
	
	\bibitem{He2013a}
    P. He, X. Ma, J. W. Zhang, H. B. Zhao, G. L\"{u}pke, Z. Shi, and S. M. Zhou, Phys. Rev. Lett. {\bf 110},  077203  (2013).
	
	\bibitem{Lee2004a}
	K. J. Lee, A. Deac, O. Redon, J. P. {Nozi\`{E}res} and B. Dieny, Nat. Mater. {\bf 3},  877  (2004).
	
	\bibitem{Stiles2007}
    M. D. Stiles, W. M. Saslow, M. J. Donahue, and A. Zangwill, Phys. Rev. B 75, 214423 (2007).
	
	\bibitem{Li2004}
	Z. Li and S. Zhang, Phys. Rev. B {\bf 70},  024417  (2004).
	
	\bibitem{Li2004a}
	Z. Li and S. Zhang, Phys. Rev. Lett. {\bf 92},  207203  (2004).
	
	\bibitem{Thiaville2005}
	A. Thiaville, Y. Nakatani, J. Miltat, and Y. Suzuki, EPL (Europhysics Letters) {\bf 69}, 990  (2005).
	
	\bibitem{Thiaville2007}
	A. Thiaville, Y. Nakatani, F. Pi\'{e}chon, J. Miltat, and T. Ono, Eur. Phys. J. B {\bf 60}, 15 (2007).
	
	\bibitem{Martinez2009}
	E. Martinez, L. Lopez-Diaz, O. Alejos, L. Torres, and M. Carpentieri, Phys. Rev. B {\bf 79}, 094430 (2009).
	
	\bibitem{Boone2010}
	C.~T. Boone and I.~N. Krivorotov, Phys. Rev. Lett. {\bf 104},  167205  (2010).
	
	\bibitem{Chureemart2011}
	P. Chureemart, R.~F.~L. Evans, and R.~W. Chantrell, Phys. Rev. B {\bf 83}, 184416  (2011).
	
	\bibitem{Iwasaki2013a}
	J. Iwasaki, M. Mochizuki, and N. Nagaosa, Nat. Nanotech. {\bf 8},  742 (2013).
	
	\bibitem{Sampaio2013}
	J. Sampaio, V. Cros, S. Rohart, A. Thiaville, and A. Fert, Nat. Nanotech. {\bf 8},  839  (2013).
	
	\bibitem{Petrovic2018}
	M.~D. Petrovi\'{c}, B.~S. Popescu, P. Plech\'a\v{c}, and B.~K. Nikoli\'{c}, Phys. Rev. Applied {\bf 10}, 054038 (2018).
	
	\bibitem{Stefanucci2013}
	G. Stefanucci and R. van Leeuwen, {\em Nonequilibrium Many-Body Theory of Quantum Systems: A Modern Introduction} (Cambridge University Press, Cambridge, 2013).
	
	\bibitem{Gaury2014}
    B. Gaury, J. Weston, M. Santin, M. Houzet, C. Groth, and X. Waintal,  Phys. Rep. {\bf 534},  1  (2014).
	
	\bibitem{Barnes2007}
	S.~E. Barnes and S. Maekawa, Phys. Rev. Lett. {\bf 98},  246601  (2007).
	
	\bibitem{Ohe2009}
	J.-I. Ohe and S. Maekawa, J. Appl. Phys. {\bf 105},  07C706 (2009).
	
	\bibitem{Shimada2015}
	Y. Shimada and J.-I. Ohe, Phys. Rev. B {\bf 91},  174437  (2015).
	
	\bibitem{Yamane2016}
    Y. Yamane, J. Ieda, and J. Sinova, Phys. Rev. B {\bf 93}, 180408(R) (2016).
    
	\bibitem{Croy2009}
	A. Croy and U. Saalmann, Phys. Rev. B {\bf 80},  245311  (2009).
	
	\bibitem{Popescu2016}
	B.~S. Popescu and A. Croy, New J. Phys. {\bf 18},  093044  (2016).
	
	\bibitem{Breuer2002}
	H.-P. Breuer and F. Petruccione, {\em The Theory of Open Quantum Systems} (Oxford University Press, Oxford, 2002).
	
	\bibitem{Zhang2004a}
	S. Zhang and Z. Li, Phys. Rev. Lett. {\bf 93},  127204  (2004).
	
	\bibitem{Fransson2010}
	J. Fransson, Phys. Rev. B {\bf 82},  180411  (2010).
	
	\bibitem{Saygun2016}
	T. Saygun, J. Bylin, H. Hammar, and J. Fransson, Nano Lett. {\bf 16},  2824 (2016).
	
	\bibitem{Taniguchi2015}
	T. Taniguchi, K.-J. Kim, T. Tono, T. Moriyama, Y. Nakatani, and T. Ono, Appl. Phys. Express {\bf 8},  073008  (2015).
	
	\bibitem{Berger1986}
	L. Berger, Phys. Rev. B {\bf 33},  1572  (1986).
	
	\bibitem{Volovik1987}
	G.~E. Volovik, J. Phys. C: Solid State Phys. {\bf 20},  L83 (1987).
	
	\bibitem{Stern1992}
	A. Stern, Phys. Rev. Lett. {\bf 68},  1022  (1992).
	
	\bibitem{Duine2008}
	R.~A. Duine, Phys. Rev. B {\bf 77},  014409  (2008).
	
	\bibitem{Tserkovnyak2008a}
	Y. Tserkovnyak and M. Mecklenburg, Phys. Rev. B {\bf 77},  134407  (2008).
	
	\bibitem{Yang2009}
	S. A. Yang, G. S. D. Beach, C. Knutson, D. Xiao, Q. Niu, M. Tsoi, and J. L. Erskine,  Phys. Rev. Lett. {\bf 102},  067201  (2009).
	
	\bibitem{Freimuth2018} 
	F. Freimuth, S. Bl\"{u}gel, and Y. Mokrousov, {\tt arXiv:1806.04782}.
	
	\bibitem{Sayad2016}
	M. Sayad, R. Rausch, and M. Potthoff, Phys. Rev. Lett. {\bf 117},  127201 (2016).
	
	\bibitem{Mondal2018a}
	P. Mondal, M. D. Petrovi\'{c}, P. Plech\'a\v{c}, and B. K. Nikoli\'{c}, {\tt arXiv:1809.09090}.
	
	\bibitem{Mahfouzi2014}
	F. Mahfouzi and B.~K. Nikoli\'{c}, Phys. Rev. B {\bf 90},  045115  (2014).
	
	\bibitem{Elze2012a}
	H.-T. Elze, Phys. Rev. A {\bf 85},  052109  (2012).
	
	\bibitem{Weston2016}
	J. Weston and X. Waintal, Phys. Rev. B {\bf 93},  134506  (2016).
	
\end{thebibliography}

\end{document}